\documentclass[11pt,a4paper]{article}

\usepackage[%
backend=biber,
style=apa,
style=authoryear,
natbib=true,
giveninits=true,
uniquename=init,
maxbibnames=6,
maxcitenames=2,
dashed=false
]{biblatex}
\usepackage{amsmath}
\usepackage{amsfonts}
\usepackage{amssymb}
\usepackage{mathtools}
\usepackage{hyperref}
\usepackage{float}
\usepackage{bbold}
\usepackage{subcaption}
\usepackage{color}
\usepackage{soul}
\usepackage{booktabs}
\usepackage{multirow}

\usepackage[title]{appendix}

\usepackage{graphicx}

\usepackage{amsthm}
\newtheorem{theorem}{Theorem}
\newtheorem{lemma}{Lemma}
\newtheorem{corollary}{Corollary}
\theoremstyle{definition}
\newtheorem{definition}{Definition}

\hypersetup{pdfauthor=author}

\newcommand{\bu}{\mathbf{u}}

\newcommand{\bm}{\mathbf{m}}
\newcommand{\bx}{\mathbf{x}}
\newcommand{\bv}{\mathbf{v}}

\newcommand{\by}{\mathbf{y}}

\newcommand{\ud}{\,\mathrm{d}}
\newcommand{\half}{{\frac 12}}
\DeclareMathOperator*{\argmax}{argmax}

\definecolor{colhlc}{cmyk}{0,0,.7,0}

\definecolor{colhlb}{cmyk}{.3,0,0,0}

\usepackage[margin=2.5cm]{geometry}

\usepackage{lmodern}
\usepackage[T1]{fontenc}

\begin{document}

\title{A Gaussian process model for chemoinformatics with application to the hazard
classification of organic solvents}
\author{Arron Gosnell\\ \texttt{arron@bowdendata.com} \\
           University of Bath, UK
           \and 
           Evangelos Evangelou\\ \texttt{ee224@bath.ac.uk}\\
           University of Bath, UK
           }

\date{}
\maketitle

\begin{abstract}
  With the proliferation of screening tools for chemical testing, it is now
  possible to create vast databases of chemicals easily. However, rigorous
  statistical methodologies employed to analyse these databases are in their
  infancy, and further development to facilitate chemical discovery is
  imperative. In this paper, we address the challenge of predicting an
  organic solvent's water pollution class based on its chemical structure. We
  hypothesise that solvents with similar chemical structure are more likely
  to belong in the same hazard class. This chemical similarity is measured by
  the Tanimoto distance, a non-Euclidean metric on the chemical space. To
  incorporate the similarity between chemical structures in the model, we
  propose a Gaussian process model on the chemical space, with the kernel
  being a function of the Tanimoto distance. A novel feature of the proposed
  model is the
  inclusion of a scaling parameter in the kernel, which controls the strength
  of the correlation between compounds and offers additional flexibility.
  We find that accounting for correlation between chemical compounds
  substantially improves predictive performance over the uncorrelated model,
  where compound similarity is unaccounted for. Our model also compares favourably
  against other established models in the literature. Furthermore, we present
  a genetic algorithm designed to identify important features to a compound's
  efficacy, with the goal of facilitating chemical discovery. The algorithm
  operates based on two criteria derived from the proposed model. Simulation
  studies are conducted to
  demonstrate the suitability of the proposed methods.\\

\end{abstract}

\begin{refsection}

\section{Introduction}\label{sec:intro}

Drug discovery and bioactivity prediction are of vital importance to many
fields, including agricultural sciences, chemistry, medicine, and the food
and drinks industry. Chemoinformatics, which focuses on the analysis of data
from chemical compounds, can aid in the understanding of influential chemical
structures and the discovery of novel drugs. Many chemoinformatics methods
rely on quantitative structure-activity relationship (QSAR) techniques to
predict bioactivities from chemical structures. To that end, chemical
representations, such as SMILES strings (a textual representation of the
chemical structure), and molecular descriptors, such as fingerprints, have
facilitated the application of a range of machine learning techniques
\citep{baptista2022evaluating, paul2018chemixnet, sandfort2020structure}.

The motivation for the paper is the classification of organic solvents
according to their hazardousness to water. Germany's water pollution
classification system (WGK) classifies chemicals as slightly (level 1),
obviously (2), or highly (3) hazardous to water, in increasing
severity. Classification is determined by accumulating evaluation points. For
instance, a solvent that is toxic if swallowed adds 3 points. The total
points are then used to assign a hazard level: 0 to 4 is level 1, 5 to 8 is
level 2, and 9 or more is level 3. Determining a solvent's hazard level prior
to manufacturing it is crucial for sustainability and cost effectiveness. It
makes sense in this context to assume that solvents with similar chemical
structure will exhibit similar properties, and are therefore more likely to
belong to the same WGK class.

Compounds are said to live within the chemical space. A central principle of
chemoinformatics is that similar compounds, i.e., compounds close in the
chemical space, share similar properties \citep{Bender2004, Dryden07}.
Closeness can be defined using metrics on dichotomous feature spaces
(fingerprints). Among such metrics, the Tanimoto similarity is the most
widely used measure, and scores highest in terms of capturing compound
similarity \citep{Bajusz2015}. \citet{Moss2020} used the Tanimoto similarity
directly as the kernel in Gaussian process regression, while
\citet{Swamidass2005} proposed a hybrid kernel based on linear combinations
of the Tanimoto similarity. A notable criticism of these methods is the
absence of a scale parameter for controlling the strength of the similarity
between compounds, thereby not properly accounting for its effect in the
model.

Motivated by the aforementioned similarity principle, in Section~\ref{GPC},
we present a novel approach to incorporating chemical distance into Gaussian
process (GP) models. The proposed GP model is defined on the chemical space,
i.e., its inputs are the chemical compounds, while the values of the GP
represent the effect of each compound on the outcome we wish to model.
Historically, GPs are defined on Euclidean spaces, owing to their early use
in geostatistics. The metrics employed for analysing chemical structures are
non-Euclidean. Replacing the Euclidean metric by some other metric
in the GP covariance does not necessarily yield a valid covariance. In
Section~\ref{sec:fp}, we provide a rigorous framework to demonstrate
that, indeed, GPs can be defined on such discrete and non-Euclidean spaces,
such as the chemical space, by incorporating the Tanimoto metric within the
GP's covariance structure. In addition, we present suitable isotropic
correlation functions adapted to live on the chemical space. An important
distinction between our proposed method and existing approaches is that we
provide the GP kernels \emph{with a scaling parameter}. To our knowledge,
this is the first paper where such kernels based on the Tanimoto metric are
developed, and an addition to the existing literature of kernels on discrete
spaces \citep[see][and references therein]{Li2025}.

We account for the fact that the hazard class is ordered through the use of a
cumulative link model with correlated random effects. Since the likelihood of
the proposed model is not available in closed form, the model is fitted using
the approximate likelihood from Laplace approximation. This
approach, detailed in Section~\ref{sec:methods}, constitutes another key
contribution of this paper: the application of the Laplace approximation
for estimation and prediction of ordinal data with Gaussian process random
effects. This extends \citet{Chu2005}
to more general link and correlation functions, that are suitable for
chemical inputs. Due to the correlation structure of the GP model, we can
gain information from the effects of sampled compounds to predict the effect
of unsampled compounds, as well as provide uncertainty values for our
predictions. These values can be incorporated within a sampling design
criterion for drug discovery, making the proposed model particularly
attractive, especially when considering the
cost-effectiveness and environmental impact of chemical production.

Exploration of the chemical space is crucial for discovering new and
effective compounds. Since the chemical space encompasses an immensely vast
number of molecular structures, it is impossible to assess all configurations
of molecular features to discover the ideal compound, making virtual
screening computationally challenging. We therefore seek efficient
optimisation techniques to automate discovery and identify chemical features
for further analysis. To that end, in Section~\ref{sec:op}, we develop a
genetic algorithm, aided by the proposed model, to search over the chemical
space and identify compounds of potentially high efficacy. We propose two
optimality criteria that can be used for this purpose that are based on the
features of the proposed model. The first criterion is based on maximising
the probability that the outcome will belong to a given class, under given
experimental conditions. Our second criterion uses an approximation of
the aforementioned probability and depends only on the parameters of the
distribution of the GP, but does not assume any experimental conditions.

Section~\ref{sec:simulation-study} presents simulation studies to demonstrate
that the proposed method recovers the true model parameter under the correct
model, and that the genetic algorithm can identify the optimal compound.
Section~\ref{sec:hazard-class-organ} applies the model to the
problem of hazard classification for organic solvents introduced
earlier.

\section{GP classification based on a cumulative probability model}\label{GPC}

We consider a chemical space $\mathcal{C} = \{c_1,\ldots,c_m\}$ of $m$
compounds. In practice, $m$ is large, but only a few compounds are used in
experiments. We assume observed data
$\{(\bx_i,y_i,c_{l_i}),\ i = 1,\ldots,n\}$, where
$y_i \in \lbrace1,2,\ldots, C \rbrace$, with $1 < 2 < \ldots < C$, is the
class response, $\bx_i \in \mathbb{R}^p$ are the testing conditions, and
$l_i \in \{1,\ldots,m\}$ is the compound indicator for the $i$th experiment.
We wish to predict the outcome $y_*$ given experimental conditions $\bx_*$
with compound $c_{*}$, i.e., to estimate the probabilities $\Pr(y_* = j|\by)$
for $j \in \{1,\ldots,C\}$, where $\by = (y_1,\ldots,y_n)$.

Let $T(\cdot,\cdot)$ denote the Tanimoto distance (defined Section~\ref{sec:fp}) between pairs of
compounds within the chemical space. We define
$u:\mathcal{C} \mapsto \mathbb{R}$ to be a GP on $\mathcal{C}$, such that
$\bu = (u(c_{1}),\ldots, u(c_{m}))$ is distributed according to the
$m$-dimensional multivariate normal distribution with mean $0$ and
variance-covariance matrix $K$. We write the $(r,s)$th element of the matrix
$K$, corresponding to compounds $c_r$ and $c_s$, where $r,s = 1,\ldots,m$, as
$k_{rs} = \sigma^2 R(T(c_{r},c_{s}),\phi)$, where $\sigma^2$ denotes the
variance parameter, and $R(t,\phi)$ denotes the correlation function at
distance $t$ with scaling parameter $\phi$. Specific forms of $R(t,\phi)$ are
derived in Section~\ref{sec:fp}.
 
Let $y$ denote the outcome of an arbitrary experiment under conditions $\bx$ with compound $c$, and let 
 $\gamma_j = \Pr(y \leq j|u(c))$, with $\gamma_C = 1$.  
Our model assumes that
\begin{equation}
  \label{eq:1}
     G(\gamma_j) = \eta_{jc} = \alpha_j + \beta^\top \bx + u(c),\ j = 1,\ldots,C-1,
\end{equation}
where $G:(0,1) \mapsto \mathbb{R}$ is the link function, ${\beta} \in \mathbb{R}^p$ are the regressor coefficients, and $\alpha_1 < \ldots < \alpha_{C-1}$ are the ordered intercepts.
Link functions model the non-linear effect of the regressor variables and the GP to the cumulative probabilities.
\citet{czado1992effect} showed that link misspecification can result in biased
estimates and higher prediction error. In practice, a suitable link function
should be chosen based on goodness-of-fit criteria, such as cross-validation.

Let $\gamma_{ij} = \Pr(y_i \leq j|u(c_{l_i}))$, $j = 1, \ldots, C$, with $\gamma_{iC} = 1$, be the cumulative probabilities for up to class $j$, and $\pi_{i1} = \gamma_{i1}$, $\pi_{ij} = \gamma_{ij} - \gamma_{i, j-1}$, $j = 2, \ldots, C$ be the individual class probabilities. 
We assume that the distribution of each $y_i$ is conditionally independent of $y_{i'}$ for $i' \neq i$ given $u(c_{l_i})$. 
Thus our model can be described by 
\begin{equation}
\label{eq:model}
     y_i | u(c_{l_i}) \stackrel{\text{ind}}{\sim}
                        \text{Categorical}(\boldsymbol{\pi}_i),\ i =
                        1,\ldots,n,\quad 
     \bu \sim \text{N}_m(0, K),
\end{equation}
where $\boldsymbol{\pi}_i = (\pi_{i1},\ldots,\pi_{iC})$ and $\bu$ is the value of the GP at the $m$ distinct compounds. 

The GP models are defined so that, if $G(\cdot)$ is increasing, low values of $u(c)$ correspond to high probabilities of an outcome in the highest class, $C$. 
To demonstrate this, we consider the odds ratio $(1-\gamma_j)/\gamma_j$, for $j=1,\ldots,C-1$, and its behaviour as a function of $u(c)$. 
We observe that  $(1-\gamma_j)/\gamma_j = 1/\gamma_j - 1 = 1/G^{-1}(\eta_{jc}) - 1$, where $\eta_{jc} = \alpha_j + \beta^\top\bx + u(c)$. 
Therefore, if $G$ is an increasing function, then so is $G^{-1}$, and in
that case, the odds ratio of observing a class higher than $j$ is a decreasing function of $u(c)$.

\section{GP on the chemical space}\label{sec:fp}

\begin{figure}
  \centering
  \includegraphics[width=.5\linewidth]{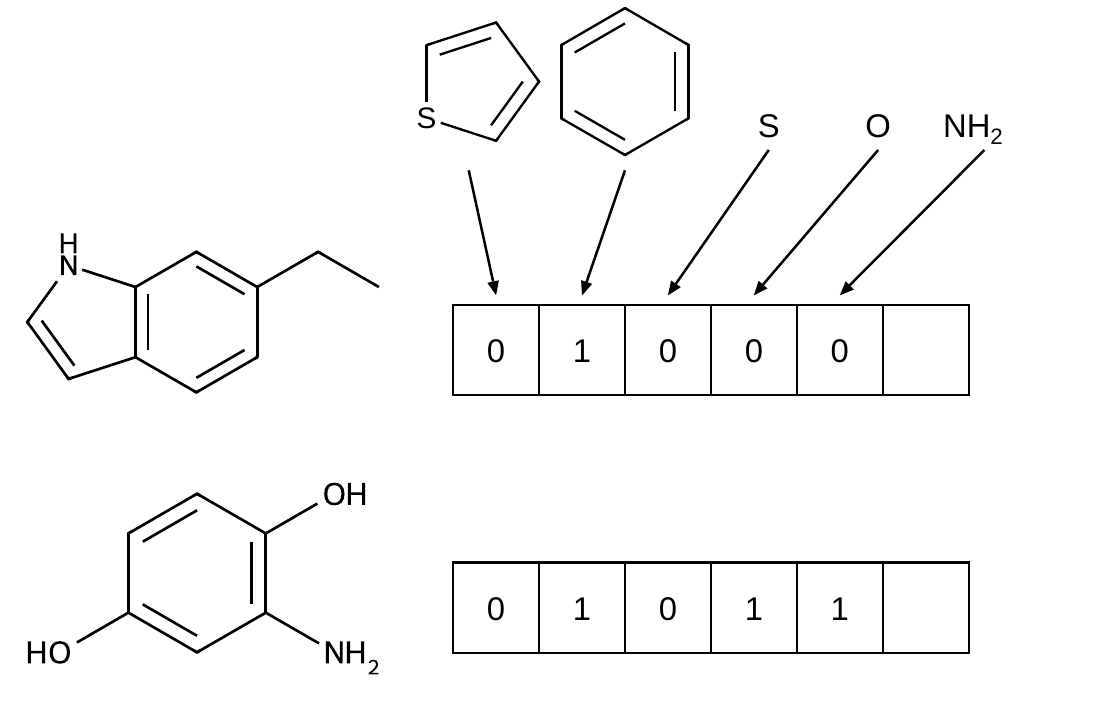}
  \caption{Two molecules and their corresponding fingerprints based on their substructures. The second feature within the two fingerprints has the value of 1, indicating the common presence of the corresponding substructure.}
\label{fpp2}
\end{figure}

Chemical fingerprints are a widely used concept in the analysis of
molecular structures and biochemical activities. Fingerprints are typically
represented as $\kappa$-dimensional bit vectors, with the features being
based on their chemical composition, or graph. Each feature within the
fingerprint indicates the presence of atomic substructures, such as
functional groups, ring systems, or atom arrangements. For example, a
fingerprint might have a bit set to 1 if a certain functional group (like a
hydroxyl group) is present within the molecule. Figure~\ref{fpp2} illustrates
two simple molecules and their associated fingerprints. %

The Tanimoto similarity captures the intersection of molecular properties
between compounds and can be used as a measure of their closeness in the
chemical space.
In defining the Tanimoto similarity, consider a collection of compounds and
their associated fingerprints
of
the form $c_{r} = (c_{r1}, c_{r2},\ldots, c_{r\kappa})$, where $c_{ri}$ is
either 0 or 1, but not all 0, denoting the presence of feature (atomic
substructure) $i$ in the $r$th compound, $i = 1,\ldots,\kappa$. The Tanimoto
similarity $S_{rs} = S(c_r,c_s)$, between the compounds $c_r$, $c_s$, is
defined to be the number of features in common between the two fingerprints over
the number of features in either. More specifically,
\begin{equation}\label{tani}
  S_{rs} = \frac{\sum_{i=1}^\kappa \min\{c_{ri},c_{si}\}}{\sum_{i=1}^\kappa \max\{c_{ri},c_{si}\}} \in [0,1].
\end{equation}
When the two compounds have no features in common, $S_{rs}=0$, and when the
compounds have identical features, $S_{rs}=1$. An important property of the
$m\times m$ matrix $S$ with elements $S_{rs}$ is that it is positive definite 
\citep{gower1971general}, hence it is a valid correlation matrix.

Subtracting the Tanimoto similarity from 1 converts it into a distance:
\begin{equation*}
  T(c_r,c_s) = T_{rs} =  1- S_{rs},
\end{equation*}
which is a metric \citep{Lipkus1999}. \citet{fenner2020privacy}
used the Tanimoto distance directly within a Gaussian spatial kernel. This
requires caution as the Tanimoto distance is non-Euclidean, and can produce
non-positive definite correlations when used this way
\citep{christakos2000norm}.

As an example, consider the chemical space $\mathcal{C} = \{c_1 = (0,1,1), c_2
= (1,0,1), c_3 = (1,1,0), c_4 = (1,1,1)\}$. The matrix of pairwise Tanimoto
distances, $T$, and the corresponding Gaussian correlation matrix $R$ with elements $R_{rs} = \exp(-T_{rs}^2)$, are given by
\begin{equation*}
T = \begin{pmatrix}
0 & 2/3 & 2/3 & 1/3 \\
  & 0   & 2/3 & 1/3 \\
  &     & 0   & 1/3 \\
  &     &     & 0
\end{pmatrix},
\quad
R = \begin{pmatrix}
1 & 0.6412& 0.6412& 0.8948 \\
  & 1     & 0.6412& 0.8948 \\
  &       & 1     & 0.8948 \\
  &       &       & 1
\end{pmatrix} \ \text{to 4 decimal points.}
\end{equation*}
Note that the distances given in $T$ cannot correspond to distances in some Euclidean space. To see this, suppose there exist points $\varepsilon_1, \ldots, \varepsilon_4$ on some Euclidean space with pairwise distances given by $T$. Then, as $T_{14} + T_{24} = T_{12}$, $T_{14} + T_{34} = T_{13}$, and $T_{24} + T_{34} = T_{23}$, the point $\varepsilon_4$ must lie simultaneously in the middle of the edges of the equilateral triangle formed by $\varepsilon_1$, $\varepsilon_2$, and $\varepsilon_3$, which is impossible. Note also that the correlation matrix $R$ is not positive definite as its lowest eigenvalue is about $-0.036$.

Next, we discuss the use of the Tanimoto distance with well-known spatial kernels.
\begin{definition}
Let $(\mathcal{C},d)$ be a metric space. The metric $d$ is called Euclidean if for any set of points $c_1,\ldots,c_m \in \mathcal{C}$, there exist $\varepsilon_1,\ldots,\varepsilon_m \in \mathbb{R}^\alpha$ ($\alpha$ depends on $m$), such that $d(c_r,c_s) = \|\varepsilon_r - \varepsilon_s\|$ for all $r,s = 1,\ldots,m$, where $\|\cdot\|$ denotes the Euclidean norm in $\mathbb{R}^\alpha$. 
In this case, we say that the points $\{c_1,\ldots,c_m\}$ can be isometrically embedded in a Euclidean space of dimension $\alpha$.
\end{definition}

The following theorem, from \citet{gower1985properties}, can be used to show that a metric is Euclidean. We denote the $m\times m$ identity matrix by $I_m$, and the $m\times m$ matrix of ones by $J_m$.
\begin{theorem}
\label{thm:1}
Let $(\mathcal{C},d)$ be a metric space. 
\begin{enumerate}
\item The metric $d$ is Euclidean if and only if, for any set of points $c_1,\ldots,c_m \in \mathcal{C}$, the $m\times m$ matrix $B = HAH$ is positive semi-definite, where $H = I_m - m^{-1} J_m$, and $A$ is the $m\times m$ matrix with elements $A_{rs} = -d(c_r,c_s)^2/2$, $r,s = 1,\ldots,m$.
\item Furthermore, let $\alpha = \mathrm{rank}(B)$. Then,  the points $\{c_1,\ldots,c_m\}$ can be isometrically embedded in a Euclidean space of dimension $\alpha$, and $\alpha$ is the lowest dimension for which this is possible.
\end{enumerate}
\end{theorem}

\begin{corollary}
  \label{thm:2}
  The chemical space $\mathcal{C} = \{c_1,\ldots,c_m\}$ with the metric
  $d(c_r,c_s) = \sqrt{T(c_r,c_s)}$ can be isometrically embedded in
  $\mathbb{R}^{m-1}$.
\end{corollary}
\begin{proof}
  The matrix $B$ in Theorem~\ref{thm:1} is
  $B = -\half H(J_m - S)H = \half HSH$, where $S$ is the $m \times m$ matrix
  with elements given by~\eqref{tani}. As $S$ is positive definite, $B$ is
  positive semi-definite and $\mathrm{rank}(B) = m-1$, therefore, the points
  $\mathcal{C}$ can be embedded in $\mathbb{R}^{m-1}$.
\end{proof}

Corollary~\ref{thm:2} allows us to create a vast catalogue of isotropic
correlation functions using the Tanimoto distance, based on the correlation
functions used in the GP literature for Euclidean spaces. Specifically,
suppose $R_e(d,\phi)$ is an isotropic correlation function on
$\mathbb{R}^{m-1}$ at distance $d$ with parameter $\phi$. Then
$R(t,\phi) = R_e(\sqrt{t},\phi)$ is a correlation function on $\mathcal{C}$.
For example, the corresponding exponential and Gaussian correlations on
$\mathcal{C}$ are given, respectively, by
\begin{equation}
  \label{eq:3}
  R_\text{Exp}(t,\phi) = \exp(-\sqrt{t}/\phi), \text{ and } R_\text{Gau}(t,\phi) = \exp(-t/\phi^2).
\end{equation}

\section{Methodology}\label{sec:methods}

Our two primary objectives are to use the available data to (i)~estimate the
model parameters, obtained through maximising the model likelihood, and
(ii)~estimate the probability of observing each class under given
experimental conditions. The model parameters are estimated first via the
maximum likelihood method. The estimates are then used to construct the
predictive distribution for the GP and the probabilities of each outcome.

The likelihood of the model, as well as the class probabilities for the given
data, can be written only as multidimensional integrals with no closed-form
expression. Techniques based on Monte-Carlo approximations of the likelihood,
such as Monte-Carlo expectation maximisation \citep{natarajan2000monte}, can
be used. However,  these methods lack computational efficiency and, given the
high dimension of the GP, alternative methods are preferred. Therefore, we
propose the use of the Laplace approximation to compute the likelihood. 

The size of the data in relation to the dimension of the GP is an important
consideration when using Laplace approximation on binary data, first examined
by \citet{Shun1995}. In particular, the sample size $n$ should increase at a
higher rate than the dimension of the GP, $m$. Theoretically speaking, $m$ is
bounded above by $2^\kappa$, where $\kappa$ denotes the number of features in
the fingerprint vector. However, in finite samples, $m$ can be comparable
with $n$, so care must be taken when using our proposed method. Furthermore,
$\kappa$ can potentially increase when more compounds are added to the database
as more features are needed to properly distinguish the compounds and ensure
a rich representation of the space.

\subsection{Estimation of model parameters}

Let $\theta = (\alpha_1,\ldots,\alpha_{C-1}, \mathbf{\beta}, \sigma^2, \phi)$ denote the model parameters. We use the symbol $f(\cdot)$ to represent the probability density/mass function of the expression in the brackets. 
Given the model in~\eqref{eq:model}, and excluding any factors that do not depend on $\theta$ or $\bu$, we have
\begin{equation}
  \label{eq:cat}
f(\by|\bu;\theta) \propto \prod_{i=1}^n \prod_{j=1}^C \pi_{ij}^{\mathbb{1}{(y_i=j)}},\quad
    f(\bu;\theta) \propto |K|^{-1/2} \exp\left(-\frac{1}{2} \bu^\top K^{-1} \bu \right),  
\end{equation}
where $\mathbb{1}{(\cdot)}$ denotes the indicator function. The likelihood, based on data $\by$, is then
\begin{equation}
\label{eq:likelihood}
L(\theta|\by) = f(\by;\theta) = \int f(\by|\bu;\theta)
f(\bu;\theta) \ud\bu .
\end{equation}

As noted earlier, the integral in~\eqref{eq:likelihood} does not have a closed-form solution, so obtaining the maximum likelihood estimates of $\theta$ by direct maximisation of the likelihood is not possible.
We use Laplace approximation to compute the likelihood, a technique which enables approximations to integrals of the form $\int e^{-g(\bu)} \ud\bu$. 
Letting $g(\bu) = -\log [f(\by|\bu;\theta)f(\bu;\theta)]$, we may express the second order Taylor expansion of $g(\bu)$ as
\begin{equation}
  g(\bu) \approx g(\hat{\bu}) + \frac{1}{2}(\bu - \hat{\bu})^\top\hat{H}(\bu - \hat{\bu}),\label{lap}
\end{equation}
where $\hat{\bu}$ denotes the point at which the function $g(\bu)$ is minimised, and $\hat{H}$ denote the Hessian matrix of $g(\bu)$ at $\hat{\bu}$.
By substituting~{\eqref{lap}} into~{\eqref{eq:likelihood}}, we obtain the approximation to the log-likelihood (up to a constant)
\begin{equation}
\label{eq:lp_approx}
    \log L(\theta|\by) \approx -g(\hat{\bu}) -\frac{1}{2}\log|\hat{H}|.
\end{equation}
Therefore, $\hat{\theta}$ may be obtained by minimising~{\eqref{eq:lp_approx}} with respect to $\theta$. Let $\mathcal{J}(\theta,\by)$ denote the negative Hessian matrix of~\eqref{eq:lp_approx}. Then, $\mathcal{J}(\hat\theta,\by)^{-1}$ is an estimate of the variance-covariance matrix of $\hat\theta$.
Furthermore, recognising that $f(\bu|\by) \propto \exp\{-g(\bu)\}$, which from~{\eqref{lap}} is proportional to a multivariate normal density, leads to the approximation
\begin{equation}\label{eq:predu}
    \bu|\by \sim {N}_m(\hat{\bu}, \hat{H}^{-1})\ \text{approximately as $n \rightarrow \infty$}.
\end{equation}
Detailed derivations are provided in Web Appendix~\ref{sec:deta-deriv-likel}.

\subsection{Estimation of class probabilities}

The approximation in~\eqref{eq:predu} enables the prediction of a class
probabilities for an untested compound, $c_*$. We begin with the
conditional distribution $u_*|\by$, where $u_* = u(c_*)$. Using the
conditional independence of $u_*$ and $\by$ given $\bu$, we have
\begin{equation*}
f(u_*|\by) = \int f(u_*|\bu) f(\bu|\by) \ud \bu 
\approx \int f(u_*|\bu) \hat f(\bu|\by) \ud \bu =: \hat{f}(u_*|\by),
\end{equation*}
so $f(u_*|\by)$ can be approximated by a normal density,
$\hat{f}(u_*|\by)$, with mean and variance
\begin{align}
  \label{eq:-mvu}
  \mathbb{E}[u_*|\by]&\approx K_*K^{-1}\hat{\bu} & \text{Var}[u_*|\by]
  &\approx K_{**}-K_*^\top K^{-1}K_* + K_*^\top K^{-1}\hat{H}^{-1}K_* K^{-1},
\end{align}
respectively, where $K_* = \text{Cov}(\bu,u_*)$, and $K_{**}= \text{Cov}(u_*,u_*)$. 

Let $y_*$ denote the outcome of a future experiment under conditions $\bx_*$
and compound $c_*$. To obtain the predicted outcome, we require the
probabilities $\Pr(y_* = j|\by)$ for $j=1,\ldots,C$.
These can be estimated as a Gaussian mixture, by
\begin{equation}
  \Pr(y_* = j|\by) \approx \int \pi_{*j} \hat{f}(u_*|\by) \ud
  u_*, \text{ where } \pi_{*j} = \Pr(y_*=j | u_*). \label{q1} 
\end{equation}
Equation~\eqref{q1} is evaluated using numerical integration. In this paper,
we use the Gauss-Hermite quadrature method \citep{elhay1987algorithm} with 21
integration points. In fact, under the probit link, the integral
in~\eqref{q1} has an analytical expression (Web Appendix~\ref{sec:pred-probit}),
however this is not the case for general link functions.

\subsection{Variance corrections to parameter uncertainty}

The formula for $\text{Var}[u_*|\by]$ given in~\eqref{eq:-mvu} is a function of the model parameters, $\theta$. 
In practice, $\theta$ is unknown and is replaced by its estimate $\hat{\theta}$, effectively assuming that the true value of $\theta$ is $\hat{\theta}$. 
This ignores the uncertainty in the value of $\theta$. 
\citet{booth1998standard} provided a correction to the prediction variance for generalised linear mixed models with \emph{independent} random effects. 
We follow a similar approach here to derive variance corrections to the GP estimates for our model.

Let $u_*$ be the true value and let $\hat{\bu}_*(\by,\theta) = \mathbb{E}[u_*|\by]$ be the prediction with known $\theta$. 
We want to assess the error $\hat{\bu}_*(\by,\hat\theta) - u_*$, where $\hat\theta$ is the maximum likelihood estimator for $\theta$.

We write $\hat{\bu}_*(\by,\hat\theta) - u_* = \hat{\bu}_*(\by,\hat\theta) - \hat{\bu}_*(\by,\theta) + \hat{\bu}_*(\by,\theta) - u_* = e_1 + e_2$, where $e_1 = \hat{\bu}_*(\by,\hat\theta) - \hat{\bu}_*(\by,\theta)$ is the additional error due to the uncertainty in $\theta$ and $e_2 = \hat{\bu}_*(\by,\theta) - u_*$ is the error had $\theta$ been known. 
Note that, $e_1$ is a function of $\by$, but not of $u_*$, and $\mathbb{E}[e_2|\by] = \hat{\bu}_*(\by,\theta) - \mathbb{E}[u_*|\by] = 0$.
Then, 
\begin{equation*}
\mathbb{E}[e_1e_2] = \mathbb{E}[\mathbb{E}[e_1e_2|\by]] = \mathbb{E}[e_1\mathbb{E}[e_2|\by]] = 0.
\end{equation*}
Furthermore,
\begin{align*}
  e_1 &= \hat{\bu}_*(\by,\hat\theta) - \hat{\bu}_*(\by,\theta) \approx \nabla_\theta \hat{\bu}_*(\by,\theta)^\top (\hat\theta - \theta) \\
  \Rightarrow \text{Var}(e_1) &\approx
  \nabla_\theta \hat{\bu}_*(\by,\theta)^\top \mathcal{I}(\theta)^{-1}
  \nabla_\theta \hat{\bu}_*(\by,\theta),
\end{align*}
where $\mathcal{I}(\theta)$ is the Fisher information matrix of $\theta$, which can be estimated by $\mathcal{J}(\hat\theta,\by)$.
Then,
\begin{align}
  \mathbb{E}[(\hat{\bu}_*(\by,\hat\theta) - u_*)^2] &= \mathbb{E}[(e_1 +
  e_2)^2] =\text{Var}(e_1+e_2) = \text{Var}(e_1) + \text{Var}(e_2) \nonumber \\
  &\approx \nabla_\theta \hat{\bu}_*(\by,\theta)^\top \mathcal{I}(\theta)^{-1}
  \nabla_\theta \hat{\bu}_*(\by,\theta) + \text{Var}[u_*|\by]. \label{eq:vc1}
\end{align}
The second term in~\eqref{eq:vc1} is given by~\eqref{eq:-mvu}, while the
first term is the variance correction due to estimation in $\theta$.
See Web Appendix~\ref{sec:deta-vari-corr} for details on how to
compute the first term in~\eqref{eq:vc1}.

\section{A genetic algorithm for drug-discovery}\label{sec:op}

A fundamental aspect of chemoinformatics is the ability to identify promising
compounds without the requirement of physical testing. Due to the expanse of
the chemical space and the high dimensionality of the molecular
representation, assessing the efficacy of every compound is 
impractical. Therefore, efficient search methods are required to guide
exploration of the chemical space and propose interesting regions for further
analysis.

In pursuit of the above objective, we propose a genetic
algorithm \citep{South93Hitch}. Genetic algorithms are effective
for feature selection \citep{katoch2021review,
  bouktif2018optimal}. The two defining elements of a genetic
algorithm are crossover and mutation. Crossover mirrors the natural
process of genetic inheritance, as it involves passing on a portion of genes
from each parent to its offspring. Mutation introduces randomness into the
population, mimicking the occasional genetic variations observed in
evolutionary cycles. Within each iteration of the algorithm, a generation of
compounds reproduce, resulting in an offspring population. The performance of
the population is evaluated through a fitness score. Features associated with
higher fitness scores are more likely to be passed down to subsequent
generations during the reproductive cycle. After a fixed number of cycles,
the features associated with the highest fitness scores form the fittest
individuals in the population.

To demonstrate the genetic algorithm, consider a population of an even number
of compounds, $k$, denoted $\{c_1,\ldots,c_k\}$, along with their
corresponding fitness, defined below, where each compound is represented by
its fingerprint vector $c_r = (c_{r1},\ldots,c_{r\kappa})$, for the $r$th
compound, $r=1,\ldots,k$. We then perform the following steps iteratively.
Each step produces an updated population, which we also denote by
$\{c_1,\ldots,c_k\}$.

\noindent\textbf{Selection:} Each compound is ranked according to the number of
  compounds whose fitness is lower than that compound's fitness. The population
  of compounds is then updated by sampling $k$ elements with replacement among
  $\{c_1,\ldots,c_k\}$, with the probability of choosing a particular
  compound, say $c_r$, being proportional to $a+b\rho_r$, where $a, b > 0$
  are chosen parameters of the algorithm, and $\rho_r$ is the rank of the
  $r$th compound.
  
\noindent\textbf{Crossover:} We form $k/2$ pairs $(c_r,c_{r+1})$ for
  $r=1,3,\ldots,k-1$. For each pair, we perform crossover with probability
  $p_c$: we sample an index $\lambda$ uniformly in $\{1,\ldots,\kappa\}$ and
  update \\
  $\bullet$ $c_{r,i} \leftarrow c_{r,i}$ and $c_{r+1,i} \leftarrow c_{r+1,i}$,
  for $i \leq \lambda$, and \\
  $\bullet$ $c_{r,i} \leftarrow c_{r+1,i}$ and $c_{r+1,i} \leftarrow c_{r,i}$,
for $i > \lambda$.

\noindent\textbf{Mutation:} For each compound, $c_r$, we perform mutation with
probability $p_m$. We sample an index $\delta$
uniformly in $\{1,\ldots,\kappa\}$ and update $c_{r\delta}
\leftarrow 1-c_{r\delta}$. 

In terms of fitness value, suppose that we are interested in identifying the
compound $c_*$ that is most likely to lead to the outcome $y_*=C$, under
experimental conditions, $\bx_*$, given the current data $\by$. According to
our model, this is achieved by the compound with the highest value of
$\Pr(y_* = C|\by)$, which is estimated by~\eqref{q1} for $j=C$. This
objective may be desirable if the experimental conditions are known. An
alternative objective is derived by approximating $\Pr(y_* = C|\by)$.
Assuming that the inverse link function $G^{-1}(\cdot)$ can be
well-approximated by a normal cumulative distribution function, Web
Appendix~\ref{sec:pred-probit} suggests that
$\Pr(y_* = C|\by) \approx 1 - \Phi\{a(\mu_* + b)/\sqrt{1+\sigma^2_*}\}$ for
some $a > 0$ and $b \in \mathbb{R}$. Assuming that $\sigma^2_*$ does not vary
significantly between compounds, the objective then becomes choosing the
compound with the lowest mean, $\mu_*$. This objective is independent of the
experimental conditions and avoids the numerical integration for computing
the class probabilities.

A crucial part of the methodology is the use of the prediction
formula~\eqref{q1} to determine a compound's fitness. It is therefore
imperative that the prediction error is low. In practice it is only possible
to gather data on few compounds which are then used to fit the model. These
compounds must be selected from a large data base in a way that they form a
representative sample of the chemical space. Approaches discussed in the
literature for this purpose include clustering, dissimilarity-based,
cell-based, and optimisation approaches \citep[see][Chapter~6 for a
review]{Leach2007}. \citet{evangelou2012optimal} studied optimality criteria
with the goal of minimising the average prediction variance that can be
applied here, while \citet{royle1998algorithm} proposed an algorithm that can
be used for dissimilarity-based selection.

\section{Simulation studies}
\label{sec:simulation-study}

We assess the performance of the proposed methods via simulation studies. The
data-generating model is given by equations~\eqref{eq:1} and~\eqref{eq:model},
with specific choices for covariates and link functions as described below.

\subsection{Estimation performance}
\label{sec:estim-perf}

First, we consider estimation of the model
parameters. The chemical space is formed by combining 5 features, producing
a total of $m=2^5-1=31$ distinct compounds (excluding the compound with
no active features). The data consist of $n=341$ experiments, where each of the
31 compounds was tested under 11 experimental conditions. Let
$y_{ik}$, where $i=1,\ldots,11$, and $k=1,\ldots,31$, denote the observed
outcome at the $i$th experiment with compound $k$, which can be among $C=3$
categories. The model for the cumulative probabilities is
\begin{equation}
  \label{eq:2}
  \mathrm{logit}\Pr(y_{ik} \leq j) = \alpha_j + \beta x_i + u_k,\ j = 1,2,
\end{equation}
with a single covariate $x_i = (i-1)/10$. We consider two different GP models
for $\bu$, one using the Gaussian, and one using the exponential
correlation (see~\eqref{eq:3}), both with variance parameter $\sigma^2$ and scale parameter
$\phi$. The model parameters $\alpha_1$, $\alpha_2$, $\beta$, $\sigma^2$, and
$\phi$ are considered unknown. We conducted two different simulation studies
from each model with the true parameter values shown in
Table~\ref{tab:sim1}. We performed 500 simulations in total from each model,
where we estimate the parameters using the proposed method.
Table~\ref{tab:sim1} shows the average estimate across the 500 simulations of
each parameter and for each model, in addition to the true and estimated
standard deviations, based on the inverse of the Hessian matrix of the
approximate log-likelihood. The results in Table~\ref{tab:sim1} show that the
proposed method can estimate the parameters accurately. In particular, for
estimating the parameters $\alpha_j$ and $\beta$, there is virtually no bias.
We estimate the standard deviation of the parameter estimates using the inverse of the
Hessian matrix of the approximate log-likelihood. We note that the standard
deviation is slightly underestimated from all parameters, except in the case of
the scale parameter $\phi$ where it is overestimated. Web
Figure~\ref{fig:simulation_box_plot} shows the distribution of
the parameter estimates from the simulation studies. It is apparent
that the estimates exhibit almost no bias (apart from the estimation of
$\phi$ is some cases), as the true value frequently lies near the median of
the estimated values. Overall, we can conclude that the proposed methodology
provides accurate estimates and standard errors for the model parameters.

\begin{table}
  \centering
  \begin{tabular}{lrcrrcrrrrrr}
    \cline{4-6} \cline{8-10}                                    
    &&&\multicolumn{3}{c}{Gaussian covariance}      &&\multicolumn{3}{c}{Exponential covariance}\\  
    \cline{1-2} \cline{4-6}                          \cline{8-10}                        
    & True    && Est & StDev & Est SD      && Est & StDev & Est StDev\\   
    \cline{1-2} \cline{4-6}                          \cline{8-10}                        
    $\alpha_1$ & $-$1.0 && $-$0.99 & 0.34 & 0.30   && $-$0.99 & 0.41 & 0.28 \\
    $\alpha_2$ &    0.0 &&    0.01 & 0.34 & 0.29   && 0.01    & 0.40 & 0.27\\
    $\beta$    &    1.0 &&    1.00 & 0.35 & 0.34   && 1.00    & 0.35 & 0.34 \\
    $\sigma^2$ &    0.5 &&    0.44 & 0.28 & 0.23   && 0.38    & 0.26 & 0.22 \\
    $\phi$     &    0.5 &&    0.49 & 0.40 & 1.31   && 0.33    & 0.53 & 2.40\\ 
    \cline{1-2} \cline{4-6}                          \cline{8-10}                        
    $\alpha_1$ & $-$0.5 && $-$0.50 & 0.29 & 0.29  && $-$0.50 & 0.29 & 0.29\\ 
    $\alpha_2$ & 0.5    && 0.50    & 0.28 & 0.29  &&   0.50  & 0.28 & 0.30\\ 
    $\beta$    & $-$1.0 && $-$0.99 & 0.36 & 0.35  && $-$0.99 & 0.36 & 0.35\\ 
    $\sigma^2$ & 1.0    && 0.96    & 0.38 & 0.38  &&   0.96  & 0.38 & 0.39\\ 
    $\phi$     & 0.1    && 0.16    & 0.22 & 0.82  &&   0.07  & 0.19 & 0.60\\ 
    \cline{1-2} \cline{4-6}                          \cline{8-10}                        
  \end{tabular}
  \caption{Performance measures for estimation of the parameters. Showing
    the true parameter values, average estimates across all simulations,
    standard deviation of the estimates across all simulations, and average
    of the standard deviation estimates across all simulations.}
  \label{tab:sim1}
\end{table}

As a comparison with alternative approaches, we consider estimation of the
class probabilities under three models: (a)~the proposed model, (b)~a
proportional odds model, and (c)~a random forest model. With models~(b)
and~(c), the 5 fingerprint features were used as additional explanatory
variables. For each model we estimate $\Pr(y_{ik} = j)$ for $i=1,\ldots,341$
and $j=1,2,3$. These estimates are compared (in terms of the Hellinger
distance) with the simulated values from~\eqref{eq:2}. We find that the
proposed model has about 50\% better performance compared to the random
forest model, and about 30\% better performance compared to the proportional
odds model. This result verifies that the proposed techniques provide
accurate estimates of the class probabilities.

\subsection{Assessment of the prediction variance formula}
\label{sec:assessm-pred-vari}

Next, we evaluate the accuracy of the variance correction
formula~(\ref{eq:vc1}). Using the simulated data from
Section~\ref{sec:estim-perf}, we predict the GP value at the 31 compounds for
each of the 500 simulated data sets. We then compute the empirical variance
for the GP corresponding to each compound across the 500 simulations and
compare it against the average of the uncorrected and corrected prediction
variance estimates. Table~\ref{tab:simvu} shows the average (over the 31
compounds) squared difference between the empirical variance and the
two estimates. We observe that the corrected version is
more accurate, and, in fact, examination of the individual estimates shows
that the uncorrected version underestimates the variance. This verifies that
the corrected prediction variance formula~(\ref{eq:vc1}) is more accurate.

\begin{table}
  \centering
  \begin{tabular}{c@{\quad}c@{\quad}c@{\quad}c@{\quad}c}
    \multicolumn{2}{c}{Gaussian} && \multicolumn{2}{c}{Exponential}\\
    \cline{1-2}\cline{4-5}
    U & C && U & C\\
    \cline{1-2}\cline{4-5}
    0.0099 &  0.0014 && 0.0029 & 0.0002 \\
    \cline{1-2}\cline{4-5}
    0.1592 &  0.1214 && 0.1557 & 0.1141\\
    \cline{1-2}\cline{4-5}
  \end{tabular}
  \caption{Average squared differences between the empirical variance and
    the uncorrected variance estimate (U) and between the empirical variance and
    the corrected variance estimate (C). The models are as in
    Table~\ref{tab:sim1}.}
  \label{tab:simvu}
\end{table}

\subsection{Assessment of the drug discovery algorithm}
\label{sec:assessm-drug-disc}

Finally, we assess the ability of the proposed
genetic algorithm of Section~\ref{sec:op} to identify compounds of high
efficacy. We simulated from the same four models as in
Section~\ref{sec:estim-perf}, except that the chemical space was increased
to 10 features, i.e. $2^{10}-1 = 1023$ compounds (excluding the compound
with no active features). In addition, only $m=90$ compounds were tested.
These compounds were selected based on a space-filling criterion using the
Euclidean distance $d(\cdot,\cdot) = \sqrt{T(\cdot,\cdot)}$
\citep{royle1998algorithm}. After fitting the model to each generated data
set, we use the proposed genetic algorithm to find the compound that
(a)~corresponds to the lowest GP value, and (b)~corresponds to the highest
probability of an output in the third category when $x=1$. The genetic
algorithm was run with population size $k=10$, for 100 generations, and
with parameters $a=10$, $b=1$, $p_c = 0.2$, and $p_m=0.9$. We applied this
method to 100 generated data sets simulated from each of the four models.
We then compared how highly ranked the derived compound is, compared to the
truly optimal compound as predicted by the fitted model, i.e., we
compute~\eqref{eq:-mvu} and~\eqref{q1} for each of the 1023 compounds in
our model, which we then rank from best to worst, and then find the rank
that corresponds to the compound selected by the genetic algorithm.
Table~\ref{tab:simga1} shows how many times each rank was attained. On
average, more than 70\% of the time, the genetic algorithm returned the
optimal compound, and more than 90\% of the time it returned one of the top
two compounds. These results can be improved by increasing the
population size $k$ and the number of iterations, with the additional
computational cost. This suggests that the proposed algorithm works well
for proposing compounds of high efficacy.

\begin{table}
  \centering
  \begin{tabular}{rcrrrrcrrrr}
    &&\multicolumn{4}{c}{Gaussian} &&
    \multicolumn{4}{c}{Exponential}\\
    \cline{3-6} \cline{8-11}
    && $1$ & $2$ & $3$ & $4+$ & & $1$ & $2$ & $3$ & $4+$\\
    \cline{1-1} \cline{3-6} \cline{8-11}
    GP           && 72 &  19 &   5 &  4 && 67 &  20 &   9 &  4 \\ 
    $\Pr(y = 3)$ && 76 &  17 &   4 &  3 && 71 &  21 &   4 &  4 \\ 
    \cline{1-1} \cline{3-6} \cline{8-11}
    GP           && 76 &  17 &   5 &  2 && 75 &  15 &   7 &  3 \\ 
    $\Pr(y = 3)$ && 74 &  15 &   6 &  5 && 71 &  20 &   4 &  5 \\ 
    \cline{1-1} \cline{3-6} \cline{8-11}
  \end{tabular}
  \caption{Frequency with which the compound returned by the genetic
    algorithm received each rank when the objective is to minimise the GP
    mean value (first and third rows), or maximise the probability in the
    third class with $x=1$ (second and fourth rows). The models are as in
    Table~\ref{tab:sim1}.}
  \label{tab:simga1}
\end{table}

\section{Hazard classification of organic solvents}
\label{sec:hazard-class-organ}

Here, we consider the list of 500 organic
solvents published in \citet{LagnerList}. The list contains a range of
chemical information, including its classification according to the German
water hazard class (WGK) discussed in Section~\ref{sec:intro}. Other
properties of the solvents are included in the data, such as GHS
classification, GHS hazard statements, Hansen solubility parameter, boiling
temperature, vapour pressure at atmospheric pressure, density, molecular
weight, and molar volume.
After removing missing values, $n=485$ data points remain in 
proportions of 25\%, 22\%, and 52\% respectively for the three WGK
classes. Of these, 343 observations (70\%) were randomly chosen as the
training set, 71 (15\%) as a validation set, and the remaining 71 as test
set. Five different training-validation-test splits
were constructed.

The goal is to predict the WGK class for each solvent. Several models were implemented: \\
{\bfseries (A) GP with Tanimoto metric:} The proposed GP ordinal model given
in~\eqref{eq:1} and~\eqref{eq:model} with Tanimoto-based correlation. The
link function was chosen among logit, probit, log-log, and complementary
log-log, and the correlation function among the Tanimoto, exponential, and
Gaussian correlations. All 12 combinations of link and correlation
functions were fitted.\\
{\bfseries (B) GP + CNN with Euclidean metric:} As (A), except that the
compounds were described by the embeddings of a convolutional neural network
(CNN). The distance between compounds was
measured by the Euclidean distance between their CNN embeddings.\\
{\bfseries (C) NN + contrastive learning:} A neural network (NN) model where
the compound similarity is learnt via contrastive learning.\\
{\bfseries (D) CNN on language model embeddings:} A 1-D CNN with inputs the
embeddings derived from the smiles strings transformed using a pre-trained language model.\\
{\bfseries (E) GP + language model embeddings:} As (B), but with
the embeddings from (D).\\
{\bfseries (F) RF:} A random forest (RF) model that is trained using the
eight other chemical properties of the solvents (not the fingerprints).\\
{\bfseries (G) XGBoost:} The XGBoost algorithm with the fingerprint features
as binary inputs.\\
{\bfseries (H) GAUCHE + Variational Bayes:} A multi-class GP classifier using
fingerprint kernels from the GAUCHE library fitted by variational Bayes methods. \\
{\bfseries (I) GLMM:} A generalised linear mixed model (GLMM) with
independent random effects.\\
Model hyperparameters were carefully selected by
minimising validation loss. Early stopping of model training by monitoring
the validation loss was also implemented. Each model was refitted three times per split
from different starting values, resulting in 15 fits per model. The test set
was used only for the final evaluation. Further mathematical and computational details of the
above models and full results are given in Web
Appendix~\ref{sec:solvents-full-results}.

The average (over the 3 runs) difference in the average logarithmic
loss over the test minus the loss of
Model~A with the logit link and Tanimoto correlation is calculated and the
average and standard deviation
over the data in the test set and over the five splits is shown in
Table~\ref{tab:solvents1} for selected models. We observe that
Models~A (the proposed model) and~H (also a GP model) are among the
best-performing models. This shows that the proposed kernels can effectively
represent the relationship between compounds. In practice, this allows the
model to  predict the probability for each WGK class for untested models, which can be helpful for
resource management. GP-based models are also faster
than neural-based models. Another advantage of the proposed model is that it
has a clear interpretation. The GLMM has significantly worse performance,
which suggests that chemical representations carry useful information.

\begin{table}
  \centering
\begin{tabular}{llrrr}
\hline
Model & Specifics & Average Loss & Std dev & Time (s)\\
\hline
\multirow{3}{*}{A}
 & logit Tanimoto & 0.000 & NA & 6.7\\
 & probit Tanimoto & 0.000 & 0.004 & 5.9\\
 & probit Gaussian & 0.001 & 0.005 & 15.2\\
\hline
\multirow{2}{*}{B}
 & probit exponential & 0.010 & 0.009 & 11.0\\
 & logit exponential & 0.013 & 0.011 & 12.1\\
\hline
C & contrastive learning & 0.016 & 0.025 & 205.8\\
\hline
D & distilled BERT & 0.014 & 0.028 & 249.5\\
\hline
E & c-log-log exponential & 0.015 & 0.032 & 254.7\\
\hline
F & Random Forest & 0.058 & 0.037 & 0.2\\
\hline
G & XGBoost & 0.021 & 0.029 & 7.6\\
\hline
\multirow{3}{*}{H}
 & Dice & 0.002 & 0.018 & 22.5\\
 & Tanimoto & 0.005 & 0.018 & 27.4\\
 & MinMax & 0.006 & 0.015 & 76.7\\
\hline
I & probit & 0.123 & 0.051 & 2.5\\
\hline
\end{tabular}

\caption{Average difference logarithmic loss, standard deviation of the
    difference in logarithmic loss, and computing time for selected models
    across the test data. The average and standard deviation are calculated
    over the data in the test set and over 5 data splits.}
  \label{tab:solvents1}
\end{table}

Next, we aim to identify the fingerprint features that contribute the most
to a solvent being classed as highly hazardous (class 3). The
proposed model with the probit link and Tanimoto correlation is fitted to
the whole data set. Only those features that are present in
more than 10\% of the solvents in our data (177 features) are considered, with the remaining
features fixed at their most common value in the data. As an exploratory
step, we find which features are more common in the class-3 solvents compared
to the other classes: when present, 3 of these features are found in more
than 80\% of the class-3 solvents, and 5 more are found in more than 70\%. We
then used the proposed genetic algorithm with population size $k=100$, number
of generations 700, and parameters $a=100$, $b=1$, $p_c=0.2$, and $p_m=0.7$
to find the solvents with the highest class-3 probability predicted
using the highest-probability criterion. Note that it is not necessary to
specify any experimental conditions as the model does not contain any
explanatory variables. We examine the features common to the 100 fittest
members from the optimisation. Twenty-six features appeared in all members of
the population, which includes the 3 most common among class-3 solvents in
the data, and 3 of the 5 that appear more than 70\% of the time. These
features could be further investigated as potential drivers of hazardousness.

\begin{figure}
  \centering
  \includegraphics[width=1\linewidth]{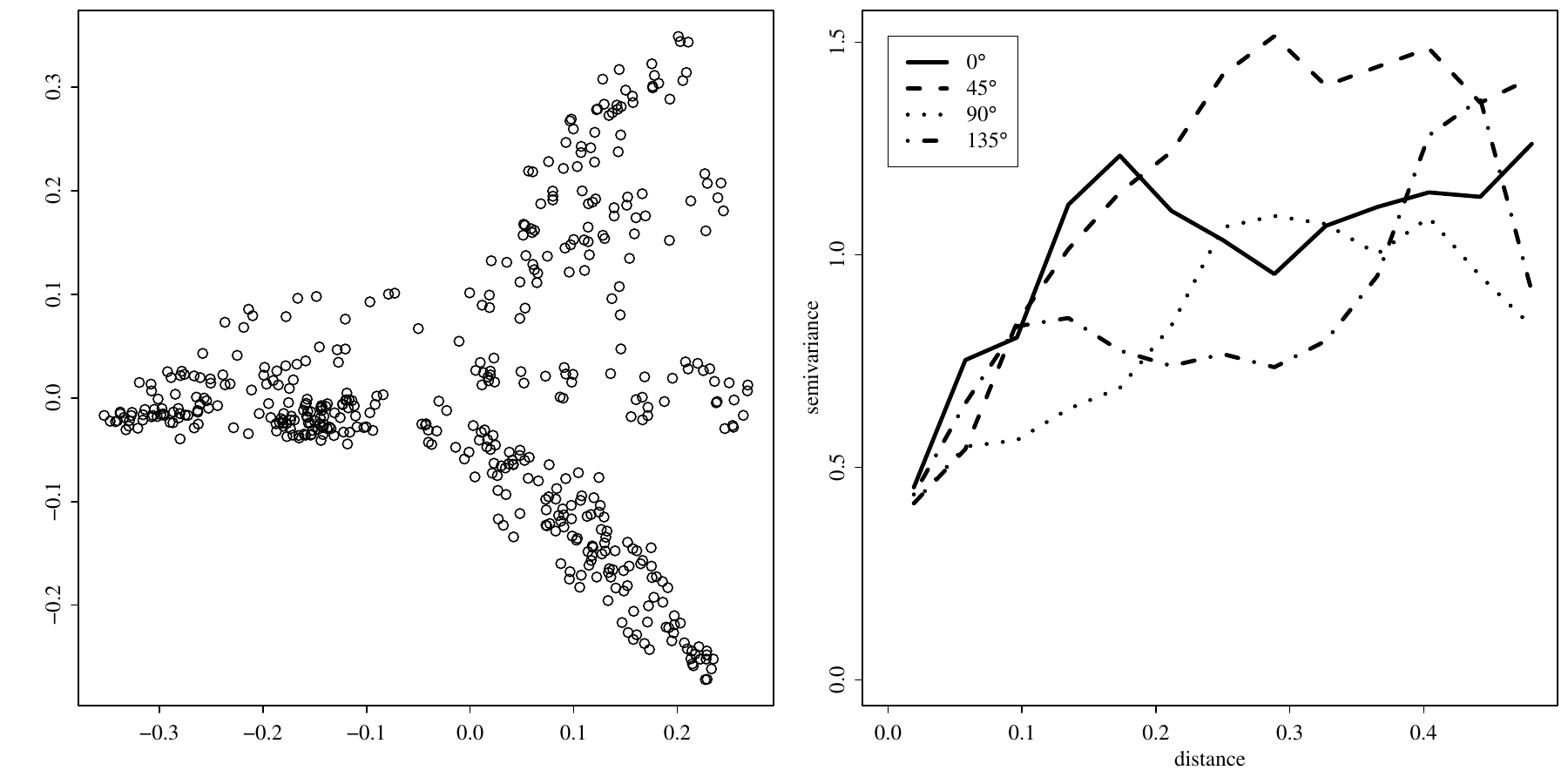}
  \caption{Left plot: Lower-dimensional embedding of the compound
    fingerprints for the data of Section~\ref{sec:hazard-class-organ}.
    Right plot: Directional semi-variogram based on the lower-dimensional
    embedding for the same data.}
  \label{fig:isotropic}
\end{figure}

An important question %
is whether the assumption of stationarity, implied by the proposed kernels,
is valid. Techniques designed to assess anisotropy in spatial data are not
directly applicable due to the high dimensionality of the embedding space. Instead, we consider a lower-dimensional embedding of the
compound fingerprints to a 2-dimensional Euclidean space using
multidimensional scaling (see Figure~\ref{fig:isotropic}, left).
Under this lower-dimensional representation, we compute (using the R package
geoR \citet{geoR}) the directional semi-variogram of the predicted GP at
various directions, which is shown in Figure~\ref{fig:isotropic} (right).
We observe the trajectories of the variograms are similar for the given
directions, suggesting the absence of anisotropy. The conclusion drawn from
this observation is that the isotropic assumption is reasonable for these
data, supporting the use of an isotropic GP model.

\section{Conclusion}

The paper is motivated by the prediction of bioactivity of chemical
compounds, and specifically the hazard classification of organic solvents.
The challenge is incorporating chemical similarity in the model, a key
principle of chemoinformatics, and utilising this towards
bio-activity prediction and drug discovery. To that end, we propose a GP
model with kernels based on the Tanimoto distance, which is a non-Euclidean
metric on the chemical space. This presents a new challenge as non-Euclidean
metrics cannot be readily used with Euclidean-based kernels, which we have
addressed through the framework of isometric embedding.

We find that the proposed GP model predicts hazard class as well as other
machine learning models, demonstrating compound similarity's relevance for
inference, yet unlike those models, ours has interpretable parameters. Our
simulations validate our techniques and show the genetic algorithm
effectively explores chemical space for identifying high-efficacy compounds.

A notable shortfall of our application is that all features within the
chemical fingerprint are considered equally important. As each fingerprint
consists of many features, examining each one individually would be
computationally expensive. A natural extension of the proposed approach is to
embed the chemical space in a higher-dimensional Euclidean space to account
for potential anisotropy or non-stationarity. Furthermore, the optimisation
method is not guaranteed to produce a realistic compound, however this issue
can be overcome by considering observable compounds that are similar to the
one derived from the optimisation. Another open question is the
identification of important fingerprint features for
prediction.
 
The proposed model may be applied to other settings, such as
predicting the potency of pharmaceutical products or the properties of food
ingredients. Directions for future work are to implement sparse correlation
functions that allow use of our methods to large chemical databases.
Further analysis could also incorporate other metrics on the chemical
space, such as the dice similarity. Another
extension would be to consider alternative representations of
the chemical space, such as those based on topological data analysis
\citep{townsend2020representation}.

\section*{Acknowledgments}

Arron Gosnell was supported by a scholarship from the EPSRC Centre for Doctoral Training in Statistical Applied Mathematics at Bath (SAMBa), under the project EP/L015684/1. Arron Gosnell acknowledges Syngenta for partial funding.

\section*{Data availability statement}

The solvents data that support the findings in this paper can be obtained from \citet{LagnerList}.

\printbibliography[title=References]

\end{refsection}

\clearpage

\begin{center}
  {\LARGE\bfseries Web-based supplementary material}\\[1em]
  {\Large Arron Gosnell and Evangelos Evangelou}
\end{center}
\vspace{1em}

\appendix

\setcounter{table}{0}
\setcounter{figure}{0}
\setcounter{equation}{0}
\setcounter{section}{0}
\setcounter{theorem}{0}
\setcounter{lemma}{0}
\renewcommand{\thepage}{\arabic{page}}
\renewcommand{\thesection}{\Alph{section}}
\renewcommand{\thetable}{S\arabic{table}}
\renewcommand{\thefigure}{S\arabic{figure}}
\renewcommand{\theequation}{S\arabic{equation}}
\renewcommand{\thelemma}{S\arabic{lemma}}
\renewcommand{\figurename}{Web Figure}
\renewcommand{\tablename}{Web Table}

\begin{refsection}
  
\section{Detailed derivations of the likelihood approximation}
\label{sec:deta-deriv-likel}

The logarithm of the probability mass function for $\by|\bu$, from~\eqref{eq:cat}, is given by
\begin{align*}
  \ell(\by|\bu;\theta) &= \sum_{i=1}^n \sum^C_{j=1} \mathbb{1}(y_i=j) \log(\pi_{ij}) \\
                       &=\sum_{i=1}^n \sum^C_{j=1} \mathbb{1}(y_i=j) \log(\gamma_{ij} - \gamma_{i,j-1})\\
                       &= \sum_{i=1}^n\sum^C_{j=1} \mathbb{1}(y_i=j) \log(G(\eta_{i,j}) - G(\eta_{i,j-1})) 
\end{align*}
where $\eta_{i,j} = \alpha_j + \beta^\top\bx + u(c_{l_i})$, and we define $\alpha_0=-\infty$, $\gamma_{i,0}=0$.
Therefore
\begin{align*}
  \frac{\partial \ell}{\partial u(c)}&= \sum_{i=1}^n \sum^C_{j=1} \mathbb{1}(y_i=j)  \mathbb{1}(c_{l_i} = c) b'_{ij}, \\
  \frac{\partial^2 \ell}{\partial u(c)\partial u(c')}&= \sum_{i=1}^n \sum^C_{j=1} \mathbb{1}(y_i=j) \mathbb{1}(c_{l_i} = c) \mathbb{1}(c_{l_i} = c') (b''_{ij} - b'_{ij}b'_{ij}),
\end{align*}
where
\begin{align*}
  b_{ij}' &= \begin{cases}
    \dfrac{G'(\eta_{i,j})}{G(\eta_{i,j})}, & \text{if $j = 1$},\\
    \dfrac{G'(\eta_{i,j})-G'(\eta_{i,j-1})}{G(\eta_{i,j})-G(\eta_{i,j-1})},
                                           & \text{if $j=2,\ldots,C-1$},\\
    \dfrac{-G'(\eta_{i,j-1})}{1-G(\eta_{i,j-1})}, & \text{if $j=C$},\\
  \end{cases} \\
  b_{ij}'' &= \begin{cases}
    \dfrac{G''(\eta_{i,j})}{G(\eta_{i,j})}, & \text{if $j = 1$},\\
    \dfrac{G''(\eta_{i,j})-G''(\eta_{i,j-1})}{G(\eta_{i,j})-G(\eta_{i,j-1})},
                                            & \text{if $j=2,\ldots,C-1$},\\
    \dfrac{-G''(\eta_{i,j-1})}{1-G(\eta_{i,j-1})}, & \text{if $j=C$}.\\
  \end{cases} 
\end{align*}

Overall, we can write 
\begin{equation*}
  \frac{\partial \ell}{\partial \bu} = P^\top \Psi_1, \qquad
  \frac{\partial^2 l}{\partial \bu \partial \bu^\top} = P^\top \Psi_2  P
\end{equation*}
where $P$ is an $n \times m$ binary matrix where its $i^{th}$ row is $0$ everywhere except at $l_i$ which equals $1$, and $\Psi_1$ is an $n$-dimensional vector and $\Psi_2$ is an $n \times n$ diagonal matrix with elements
\begin{align*}
  \Psi_{1i} &= \sum_{j=1}^C \mathbb{1}(y_i=j) b'_{ij},\\
  \Psi_{2ii} &= \sum_{j=1}^C \mathbb{1}(y_i=j) (b''_{ij} - b'_{ij}b'_{ij}),
\end{align*}
respectively, for $i=1,\ldots,n$.
To find $\hat{\bu}$ used in the Laplace approximation, we solve 
\begin{equation}
  \label{eq:usolve}
  K^{-1}\hat\bu-P^\top\hat\Psi_1=0,
\end{equation}
and the Hessian is $\hat{{H}}=K^{-1}-P^\top\hat\Psi_2P$, where $\hat\Psi_1$ and $\hat\Psi_2$ denote $\Psi_1$ and $\Psi_2$ evaluated at $\hat\bu$.

\section{Details on variance correction}
\label{sec:deta-vari-corr}

To compute the derivatives $\nabla_\theta \hat{\bu}_*(\by,\theta)$ needed
in~\eqref{eq:vc1}, note that, by~\eqref{eq:-mvu},
\begin{align*}
  \nabla_\theta \hat{\bu}_*(\by,\theta) & = \nabla_\theta (K_*K^{-1} 
                                          \hat{\bu}(\by,\theta)) \\
                                        &= \{(\nabla_\theta K_*) K^{-1} - K_*K^{-1}(\nabla_\theta K) K^{-1}\}
                                          \hat{\bu}(\by,\theta) +
                                          K_*K^{-1} (\nabla_\theta \hat{\bu}(\by,\theta)),
\end{align*}
where $\hat{\bu}(\by,\theta)$ is the solution
to~\eqref{eq:usolve}. By differentiating both sides of~\eqref{eq:usolve} with
respect to elements of $\theta$, we are able to compute
$\nabla_\theta \hat{\bu}(\by,\theta)$ algebraically.

\section{Further simulation results}

\begin{figure}[tb]
  \begin{subfigure}[t]{.48\textwidth}
    \centering
    Gaussian\par\medskip
    \includegraphics[width=\linewidth]{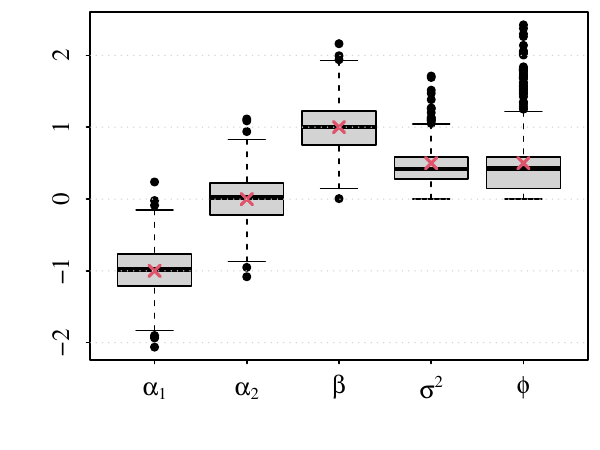}
    \caption{}
  \end{subfigure}
  \hfill
  \begin{subfigure}[t]{.48\textwidth}
    \centering
    Exponential\par\medskip
    \includegraphics[width=\linewidth]{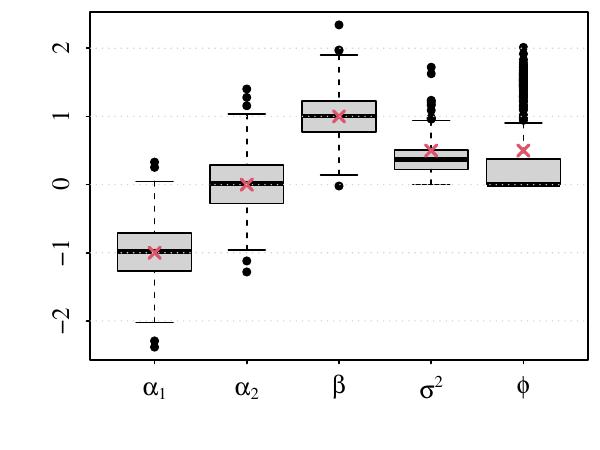}
    \caption{}
  \end{subfigure}

  \medskip

  \begin{subfigure}[t]{.48\textwidth}
    \centering
    \includegraphics[width=\linewidth]{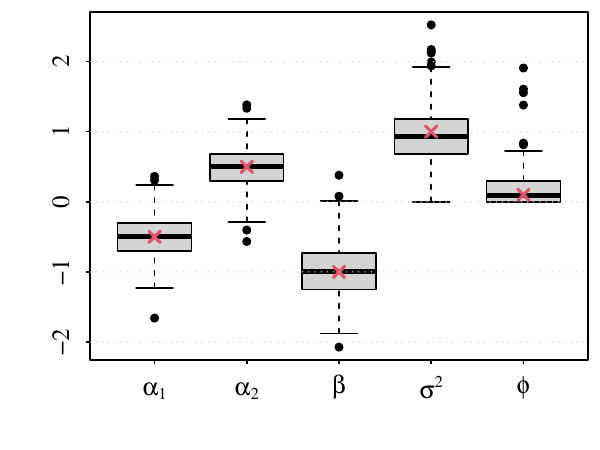}
    \caption{}
  \end{subfigure}
  \hfill
  \begin{subfigure}[t]{.48\textwidth}
    \centering
    \includegraphics[width=\linewidth]{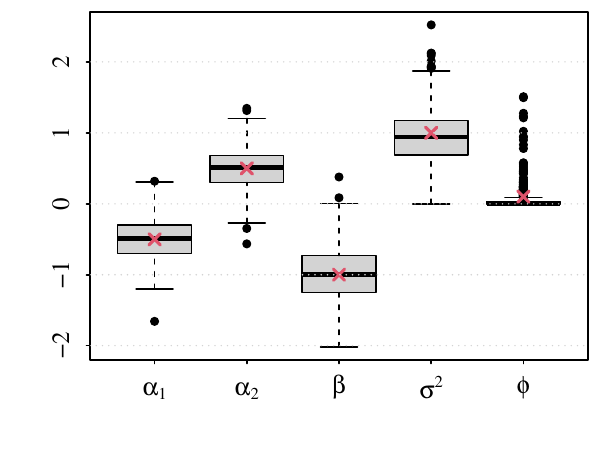}
    \caption{}
  \end{subfigure}
  
  \caption{Showing the distribution of the estimated parameter values from
    the simulation studies. The true model parameters are indicated with a red $\times$. The models are as in Table~\ref{tab:sim1}. 
  }
  \label{fig:simulation_box_plot}
\end{figure}

Web Figure~\ref{fig:simulation_box_plot} shows the distribution across
simulations of the estimates using the proposed Laplace approximation method
for each parameter for the different models considered.
\label{sec:furth-simul-results}

\section{Prediction under the probit link}
\label{sec:pred-probit}

In this section we derive the closed-form expression for the prediction probabilities of a future experiment under the probit link. These probabilities can be expressed in integral form in general by the right-hand side of~\eqref{q1}.

We will make use of the following lemma, which is proven in Section~3.9 of \citet{rasmussen2006gaussian}.
\begin{lemma}
  Let $f(z|\mu,\sigma^2)$ denote the Gaussian probability density function with mean $\mu$ and variance $\sigma^2$, and let $\Phi(z)$ denote the standard normal cumulative distribution function. Then, for $a \in \mathbb{R}$,
  \begin{equation*}
    \label{eq:gplemma}
    \int \Phi(z-a) f(z|\mu,\sigma^2) \ud z = \Phi \left(\frac{\mu-a}{\sqrt{1+\sigma^2}}\right) .
  \end{equation*}
\end{lemma}

Considering the probit model, we have for the $j$th category, $j=1,\ldots,C$,
$\pi_{*j} = \Phi(\alpha_j + \bx_*^\top\beta + u_*) - \Phi(\alpha_{j-1} +
\bx_*^\top\beta + u_*)$, with the convention $\alpha_0 = -\infty$ and
$\alpha_C= \infty$, and $\hat f(u_*|\by)$ corresponds to the Gaussian density
with mean $\mu_* = \mathbb{E}[u_*|\by]$ and variance
$\sigma^2_* = \mathrm{Var}[u_*|\by]$, given by~\eqref{eq:-mvu}. Then,
by~\eqref{q1},
\begin{align*}
  \Pr(y_* = j|\by) &\approx \int \pi_{*j} \hat f(u_*|\by) \ud u_* \\
                   &= \int \Phi(\alpha_j + \bx_*^\top\beta + u_*) \hat f(u_*|\by) \ud u_*
                     - \int \Phi(\alpha_{j-1} + \bx_*^\top\beta + u_*) \hat f(u_*|\by) \ud u_*
  \\
                   &= \Phi \left(\frac{\mu_* + \alpha_j +
                     \bx_*^\top\beta}{\sqrt{1+\sigma^2_*}} \right) -
                     \Phi \left(\frac{\mu_* + \alpha_{j-1} +
                     \bx_*^\top\beta}{\sqrt{1+\sigma^2_*}} \right).
\end{align*}

\section{Models and results of the analysis of the solvents data}
\label{sec:solvents-full-results}

We give detailed descriptions of the various models and algorithms used to
analyse the solvents data. The neural network models detailed below were
implemented using the PyTorch Python library \citep{paszke2019pytorch}. The
GP models proposed in this paper were implemented by ourselves using the
software R \citep{R} with heavier computations in Fortran~90.

Each smiles representation in the data set was first converted to a
fingerprint of size 2048 using RDKit \citep{RDKit} with the maximum number of
bonds (path length) included when hashing linear and branched molecular
substructures set to 7. The 485 data points were randomly split into 343
training, 71 validation, and 71 test subsets such that the proportions of
each class in these subsets are approximately equal to their proportions in
the whole dataset, i.e., 25\%, 22\%, and 52\% respectively. Five different
such splits were generated randomly with each model described below fitted
only using the data in the training set. The validation set was used to
select optimal hyperparameter values and to stop training early. The test set
was used to evaluate model performance. To avoid any data leakage, separate
data loaders for each subset were constructed and only the training and
validation data loaders were inputted to the training algorithm. The test
data loader was only used for the calculation of the loss criteria given in
Web Section~\ref{sec:results}. To confirm that, the test set
on one hand, and the training and validation sets on the other, are not
highly similar, the nearest neighbour Tanimoto distance for every point in
the test set was computed over the combined training and validation sets, and
the average over all points in the test set is taken, i.e.,
\begin{equation*}
  \bar{d}_\mathrm{NN} = \frac{1}{|\text{test}|} \sum_{j \in \text{test}}
  \min_{i \in \{\text{train}, \text{validation}\}} T_{ij},
\end{equation*}
where $T_{ij}$ denotes the Tanimoto distance between the $i$th compound in
the combined train and validation subset, and $j$th compound in the test
subset. The average nearest neighbour distances for splits 1 through 5 are,
respectively, 0.35, 0.33, 0.39, 0.33, 0.34.

The code that implements the methods below is available in the supplementary
material. All tuning parameters were chosen to optimise predictive
performance and reduce overfitting. For neural network models, Bayesian
optimisation as implemented by the Optuna package \citep{optuna_2019} was
used to select optimal hyperparameter values. Neural network training was
performed using either the adaptive moment estimation (adam) optimiser
\citep{kingma2015adam}, or the adam with weight decay (adamw).

All computations were performed on a laptop equipped with a 12th-generation
Intel Core i7-1260P processor, 32 GB of RAM, running Linux Mint 22.3. The
software versions used are Python version 3.12.3, R version 4.3.3,
gcc/gfortran version 13.3.0. Specific Python package versions used are gauche
0.1.6, gpytorch 1.15.2, numpy 1.26.4, optuna 3.5.0, pandas 2.1.4+dfsg, torch
2.6.0+cu118, transformers 5.12.1. Specific R package versions used are
randomForest 4.7-1.1, xgboost 1.7.11.1.

\subsection{Model A: GP with Tanimoto-based kernel}
\label{sec:MA}

This is the model proposed in this paper. We consider the proposed GP ordinal
model, with a combination of correlation and link functions as shown in
Table~\ref{tab:loss_results}. Specifically, the link functions are
\begin{align*}
  G_\text{logit}(x) &= \log \dfrac{x}{1-x},& G_\text{probit}(x)
  & = \Phi^{-1}(x),\\
  G_\text{log-log}(x) &= -\log\{-\log(x)\},& G_\text{c-log-log}(x)
  &= \log\{-\log(1-x)\},
\end{align*}
and the correlation functions are
\begin{align*}
  R_\text{Exp}(t,\phi) &= \exp(-\sqrt{t}/\phi),
  &R_\text{Gau}(t,\phi) &= \exp(-t/\phi^2),
  &R_\text{Tan}(t,\phi) &= 1-t,
\end{align*}
where $t$ denotes the Tanimoto distance.

The model parameters are estimated by the
approximate maximum likelihood method and the probabilities for each class
are calculated via numerical integration as discussed in the main paper.

\subsection{Model B: GP with Euclidean-based kernel derived from a convolutional
  neural network}
\label{sec:MB}

This model is implemented in two stages. In the first stage, as convolutional
neural network (CNN) is fitted to the training data. The CNN's embeddings
(linear predictors) for the training and test sets are then calculated and
used as a representation of the chemical space. In the second stage, a GP
model as the one described in this paper is fitted to the training data, with
the difference that the GP kernel is a Euclidean-based kernel over the CNN's
embeddings.

The following components define our CNN implementation.
\begin{enumerate}
\item \textbf{Input layer:} The input layer consists of a tensor $X^{(0)}$ of
  dimension $r \times 2048 \times 1$, where $r$ denotes the batch size,
  containing the fingerprints. Thus $X^{(0)}_{i,k,1} = c_{i,k}$ for the $i$th
  compound in the batch and the $k$th fingerprint feature.
\item \textbf{Hidden layers:} The CNN consists of $J=2$ hidden layers. Layer
  $j \in \{1,2\}$ takes as input a tensor $X^{(j-1)}$ of dimension
  $r \times l_{j-1} \times m_{j-1}$, where $l_{j-1}$ is the
  ``image'' length, and $m_{j-1}$ is the ``channel'' size, and returns a tensor
  $X^{(j)}$ of dimension $r \times l_{j} \times m_{j}$, by following
  the steps below.
  \begin{enumerate}
  \item \textbf{Convolution:} Let $W^{(j)}$ be a weight tensor of dimension
    $m_{j} \times m_{j-1} \times s_j$ where $m_{j}$ denotes the number of
    filters and $s_j$ denotes the window size, and let $B^{(j)}$ be a vector
    of dimension $m_{j}$ denoting the bias terms. The convolution of
    $X^{(j-1)}$ with $W^{(j)}$ with bias $B^{(j)}$ is a tensor $Z^{(j)}$ of
    dimension $r \times l_{j-1} \times m_{j}$ defined as
    \begin{equation*}
      Z^{(j)}_{tuv} = \sum_{i=1}^{m_{j-1}} \sum_{k=1}^{s_j} W^{(j)}_{v,i,k}
      X^{(j-1)}_{t,u+k-1,i} + B^{(j)}_{v},
    \end{equation*}
    $t = 1,\ldots,r$, $u = 1,\ldots, l_{j-1}$, $v = 1,\ldots,m_{j}$. When
    $u+k-1 > l_{j-1}$, $X^{(j-1)}_{t,u+k-1,i} = 0$.
  \item \textbf{Activation:} We compute the tensor $U^{(j)}$
    of dimension $r \times l_{j-1} \times m_{j}$ by
    \begin{equation*}
      U^{(j)}_{tuv} = \max\{0, Z^{(j)}_{tuv}\},
    \end{equation*}
    $t = 1,\ldots,r$, $u = 1,\ldots, l_{j-1}$, $v = 1,\ldots,m_{j}$. Thus the
    rectified linear unit (ReLU) activation function is used.
  \item \textbf{Max pooling:} We apply max pooling with window size $p_j$
    of the output from the previous step produces a tensor $V^{(j)}$ of
    dimension $r \times l_{j} \times m_{j}$ where
    $l_{j} = \lfloor l_{j-1}/p_j \rfloor$, i.e.,
    \begin{equation*}
      V^{(j)}_{tuv} = \max_{i = 1,\ldots,p_j}\{U^{(j)}_{t,(u-1)p_j+i,v}\},
    \end{equation*}
    $t = 1,\ldots,r$, $u = 1,\ldots, l_{j}$, $v = 1,\ldots,m_{j}$.
  \item \textbf{Dropout:} Finally, we apply a random dropout where each
    component of $V^{(j)}$ is randomly set to 0 with probability $p_d$, i.e.,
    we compute $X^{(j)}$ where
    \begin{equation*}
      X^{(j)}_{tuv} = V^{(j)}_{tuv} \times \mathbb{1}(\Upsilon_{tuv} > p_d),
    \end{equation*}
    where $\Upsilon_{tuv}$ are independent uniformly distributed random
    variables in $(0,1)$.
  \end{enumerate}

  Note that the output from the final layer, $X_{tuv}^{(J)}$, corresponds to
  the CNN embeddings.
\item \textbf{Output layer:} For classification with $C$ output classes, we
  compute:
  \begin{enumerate}
  \item The logit terms for class $y \in \{1,\ldots,C\}$ by
    \begin{equation*}
      \tilde Z^{(h)}_{t} = \sum_{i=1}^{l_{J}} \sum_{k=1}^{m_{J}} \tilde
      W^{(h)}_{ik} X^{(J)}_{tik} + \tilde{B}^{(h)},
    \end{equation*}
    for $t=1,\ldots,r$, where $\tilde{W}^{(h)}$ and is a
    $l_{J} \times m_{J}$ tensor of weights and $\tilde B^{(h)}$ is the
    bias for class $h$.
  \item The probability for class $h \in \{1,\ldots,C\}$ by
    \begin{equation*}
      P^{(h)}_t = \mathrm{Softmax}_h(\tilde Z^{(1)}_t,\ldots,\tilde Z^{(C)}_t)
      = \frac{\exp\{\tilde Z^{(h)}_t\}}{\exp\{\tilde Z^{(1)}_t\} +
        \ldots + \exp\{\tilde Z^{(C)}_t\}},
    \end{equation*}
    for $t=1,\ldots,r$.
  \end{enumerate}
\item \textbf{Loss function:} We use the cross-entropy (logarithmic) loss
  \begin{equation*}
    L_{\text{class}} = -\frac{1}{r} \sum_{t=1}^r \log P_t^{(y_t)},
  \end{equation*}
  where $y_t$ denotes the observed toxicity class.
\end{enumerate}

\subsubsection{Implementation details}
\label{sec:MB-impl}

Here, we summarise the specific training protocol and hyperparameter choices.

\begin{itemize}
\item \textbf{Input Feature Encoding:} Molecular fingerprints are passed directly as $2048$-dimensional binary vectors converted to tensors ($l_0 = 2048$, $m_0 = 1$) without tokenisation or pretraining.

\item \textbf{Hyperparameter Configuration:} The architectural parameters used in our experiments are $m_1 = 8$ and $m_2 = 16$ (filter channels), $s_1 = s_2 = 3$ (kernel sizes), $p_1 = p_2 = 2$ (max pooling windows), and $p_d = 0.5$ (dropout rate).

\item \textbf{Embedding Dimension:} The flattened CNN representation
  $X^{(J)}$ yields an embedding dimension of $l_2 \times m_2 = 512 \times 16
  = 8192$. 

\item \textbf{Optimisation Protocol:} Parameters are optimised
  adam optimiser with an initial learning rate $\eta = 3 \times 10^{-4}$, weight decay of $1 \times 10^{-3}$, and maximum gradient norm clipping of $1.0$. A scheduler reduces the learning rate by a factor of $0.5$ if validation loss plateaus for 5 consecutive epochs.
 
\item \textbf{Training and early stopping:} Training uses randomly shuffled
  mini-batches of size $r = 32$. Models are trained for up to 100 epochs with
  early stopping triggered if the logarithmic loss over the validation set
  fails to improve by $0.005$ over 3 consecutive epochs. Once early stopping
  criteria were met, the model weights corresponding to the epoch with the
  lowest validation loss were restored for out-of-sample test evaluation.
  
\item \textbf{Hyperparameter Tuning:} Hyperparameter search space is given in
  Web Table~\ref{tab:1d_cnn_params}. Hyperparameters 
  were selected via 50 Optuna optimisation trials on minimising the
  logarithmic loss over the validation subset.

\item \textbf{Replications:} Each model training procedure was run 3 times
  for each split, with different parameter starting values each time 
  to ensure robustness of the training procedure.

\item \textbf{Compute Budget:} Full
  training completes in under 1 minutes per run on a single CPU.
\end{itemize}

\begin{table}[htbp]
\centering
\begin{tabular}{lll}
\hline
\textbf{Hyperparameter} & \textbf{Values} & \textbf{Status} \\
\hline
1st layer number of channels ($m_1$) & $\{4, \mathbf{8}, 16\}$ & Tuned \\
  1st layer kernel size ($s_1$) & $\{\mathbf{3}, 5, 7\}$ & Tuned \\
1st layer pooling window ($p_1$) & 2 & Fixed \\
2nd layer number of channels ($m_2$) & $\{8, \mathbf{16}, 32\}$ & Tuned \\
2nd layer kernel size ($s_2$) & $\{\mathbf{3}, 5\}$ & Tuned \\
2nd layer pooling window ($p_2$) & 2 & Fixed \\
Dropout rate ($p_d$) & $\{0.3, 0.4, \mathbf{0.5}, 0.6, 0.7\}$ & Tuned \\
Learning rate & $\{1 \times 10^{-4}, \mathbf{3 \times 10^{-4}}, 5 \times 10^{-4}\}$ & Tuned \\
Weight decay & $\{1 \times 10^{-4}, \mathbf{1 \times 10^{-3}}, 1 \times 10^{-2}\}$ & Tuned \\
Batch size ($r$) & $\{16, \mathbf{32}, 64\}$ & Tuned \\
Gradient clip threshold & $1.0$ & Fixed \\
\hline
\end{tabular}
\caption{Hyperparameter search space for Model~B tuned via Optuna. Optimal
  values in bold.}
\label{tab:1d_cnn_params}
\end{table}

\subsubsection{Training monitoring}
\label{sec:MB-monitoring}

\begin{figure}
  \centering
  \includegraphics[width=1\linewidth]{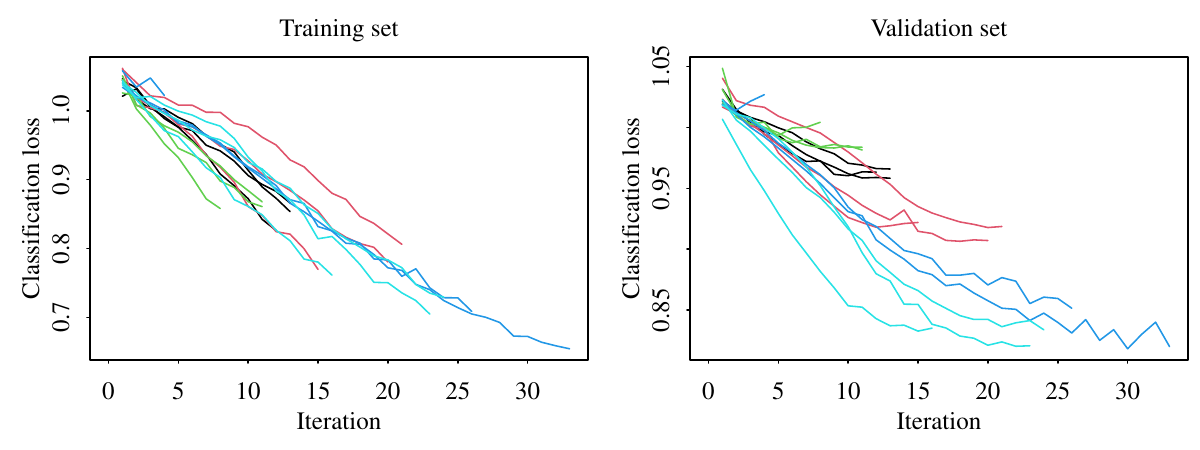}
  \caption{Training convergence curves for Model~B. The left plot shows the
    classification loss on the training set and the right plot shows the
    classification loss for the validation set against training iteration.
    Each colour corresponds to one of the data split; the three curves of the
    same colour are the three independent runs for that split. Training stops
    when the early stopping criterion is met.}
  \label{fig:ModelB_conv}
\end{figure}

Web Figure~\ref{fig:ModelB_conv} shows the classification loss for the
training and validation sets for the five different splits (different
colours) for each of the three runs for each split.

\subsection{Model C: Neural network model using contrastive learning and triplet loss}
\label{sec:MC}

This model is a multilayer perceptron (MLP) trained in two stages. In the
first stage, we use contrastive learning with triplet loss to construct an
embedding feature space where compounds sharing the same toxicity class are
close \citep{pinheiro2022smiclr}. In the second stage, the network is
fine-tuned to classify compounds using class labels. A baseline MLP was also
fitted that is only trained at the second stage (see implementation details
below). 

The network components are constructed as follows.
\begin{enumerate}
\item \textbf{Input layer:} An input tensor $X^{(0)}$ of dimension
  $r \times m_0$, where $r$ denotes the batch size, containing the binary
  molecular fingerprints. Thus $X^{(0)}_{i,k,1} = c_{i,k}$ for the $i$th
  compound in the batch and the $k$th fingerprint feature.
\item \textbf{Hidden layers:} The feature encoder consists of $J=2$ hidden
  linear transformations:
  \begin{enumerate}
  \item \textbf{First Layer \& Activation:} $X^{(1)} = \max\left\{0,X^{(0)} W^{(1)} + B^{(1)}\right\}$, where $W^{(1)} \in \mathbb{R}^{m_0 \times m_1}$ and $B^{(1)} \in \mathbb{R}^{m_1}$. Dropout with rate $p_d$ is applied to $X^{(1)}$.
  \item \textbf{Embedding Layer:} $X^{(2)} = X^{(1)} W^{(2)} + B^{(2)}$, where $W^{(2)} \in \mathbb{R}^{m_1 \times m_2}$ and $B^{(2)} \in \mathbb{R}^{m_2}$. In the contrastive stage, output embeddings are $L_2$-normalised: $\hat{X}^{(2)} = X^{(2)} / \|X^{(2)}\|_2$.
  \end{enumerate}

\item \textbf{Output layer:} The logit terms for the $C$ classes,
  $\tilde{Z}_t \in \mathbb{R}^C$, and the individual class probabilities
  $P^{(h)}_t$ are computed via a linear layer applied to $X^{(2)}$
  \begin{equation*}
    \tilde{Z}_t = X^{(2)}_t \tilde{W} + \tilde{B}, \quad P^{(h)}_t = \frac{\exp(\tilde{Z}_{t,h})}{\sum_{k=1}^C \exp(\tilde{Z}_{t,k})},
  \end{equation*}
  for $t=1,\ldots,r$, where $\tilde{W}$ is a
  $m_{J} \times C$ tensor of weights and $\tilde B$ is the $C$-dimensional
  bias term.
\end{enumerate}

\subsubsection{Implementation details}
\label{sec:MC-impl}

The tensors $W^{(j)}$ and $\tilde{W}$, and biases $B^{(j)}$ and
$\tilde B$ are considered unknown parameters and updated via the
back-propagation algorithm in two stages.

\begin{itemize}
\item \textbf{Input Feature Encoding:} Molecular fingerprints are passed
  directly as $2048$-dimensional binary vectors converted to tensors ($m_0 =
  2048$) without tokenisation or pretraining. 
\item \textbf{Architecture Hyperparameters:} Hidden dimension $m_1 = 128$, embedding dimension $m_2 = 64$, and dropout rate $p_d = 0.5$.
\item \textbf{Stage 1 (Contrastive Pretraining):}
  \begin{itemize}
  \item \textbf{Triplet Strategy:} Each training compound generates $s = 3$
    triplets per epoch containing an anchor, a positive sample (same class),
    and a negative sample (different class) chosen randomly without
    replacement. A training triplet dataset was constructed using only data
    from the training subset, and a validation triplet dataset was
    constructed using only data from the validation subset.
    To prevent overfitting to static pairs, the training triplet dataset
    re-samples new positive and negative compound pairings randomly for every
    anchor at the start of each epoch. 
  \item \textbf{Optimisation:} Optimised with Triplet Margin Loss
    using a margin parameter $E = 0.05$. We use the adamw optimiser with
    initial learning rate $1 \times 10^{-5}$, weight decay
    $1 \times 10^{-5}$, linear warmup for 5 epochs, and gradient clipping at
    threshold $1.0$.
  \item \textbf{Training and early stopping:}
    Training uses randomly
    shuffled mini-batches of size $r = 32$.
    Trained
    for up to 250 epochs with early stopping (patience of 40 epochs on
    validation triplet loss). Once early stopping criteria were met, the
    model weights corresponding to the epoch with the lowest validation loss
    were restored.
  \end{itemize}
\item \textbf{Stage 2 (Classification Fine-Tuning):}
  \begin{itemize}
  \item \textbf{Optimisation:} Optimised using Cross-Entropy Loss
    with label smoothing ($\alpha = 0.1$) using adamw. Differential learning
    rates are used: $\text{lr}_f = 5 \times 10^{-6}$ for the feature layers
    and $\text{lr}_c = 5 \times 10^{-4}$ for the classification head. The
    gradient clipping threshold was set to 1.0 and weight
    decay is set to $0.001$ for both.
  \item \textbf{Training and early stopping:}
    Training uses randomly
    shuffled mini-batches of size $r = 32$.
    Trained for up to 150 epochs. An optimisation scheduler reduces learning
    rates by $0.5$ after 5 stagnant epochs. Early stopping triggers if
    validation loss fails to improve after 15 epochs. Once early stopping
    criteria were met, the model weights corresponding to the epoch with the
    lowest validation loss were restored for out-of-sample test evaluation.
  \end{itemize}
\item \textbf{Baseline Model:} The baseline MLP (without contrastive
  pretraining) uses identical hidden dimensions ($128 \to 64$) and is
  trained directly in Stage 2 with $\text{lr}_f = 1 \times 10^{-4}$,
  $\text{lr}_c = 1 \times 10^{-3}$, and early stopping patience of 5
  epochs.
\item \textbf{Hyperparameter Tuning:}
  Stage 1 was tuned first and the optimal hyperparameter values from Stage~1
  was fixed when choosing hyperparameters for Stage~2. Hyperparameter search
  space is given in Web Table~\ref{tab:optuna_triplet}.
  \begin{itemize}
  \item \textbf{Stage 1 (Triplet Embeddings):} Optuna (50 trial runs), based
    on minimising the triplet loss over the validation set.
  \item \textbf{Stage 2 (Classification):} Optuna (50 trial runs), based on
    minimising the logarithmic loss over the validation set.
  \end{itemize}
\item \textbf{Replications:} Each model training procedure was run 3 times
  for each split, with different parameter starting values each time 
  to ensure robustness of the training procedure.
\item \textbf{Compute Budget:} Full
  two-stage training completes in under 4 minutes per run on a single CPU.
\end{itemize}

\begin{table}[htbp]
\centering
\begin{tabular}{llll}
\hline
\textbf{Stage} & \textbf{Hyperparameter} & \textbf{Values} & \textbf{Status} \\
\hline
1 & Embedding dimension ($m_2$) & $\{\mathbf{64}, 128, 256\}$ & Tuned\\
1 & Hidden layer dimension ($m_1$) & $\{64, \mathbf{128}, 256\}$  & Tuned \\
1 & Dropout probability ($p_d$) & $\{0.1, 0.3, \mathbf{0.5}\}$ & Tuned\\
1 & Learning rate & $\{\mathbf{1 \times 10^{-5}}, 5 \times 10^{-5}, 1 \times 10^{-4}\}$ & Tuned \\
1 & Weight decay & $\{\mathbf{1 \times 10^{-5}}, 1 \times 10^{-4}, 1 \times
                   10^{-3}\}$ & Tuned \\
1 & Gradient clipping threshold & 1.0 & Fixed \\
1  & Batch size ($r$) & 32 & Fixed \\
2 & Learning rate features ($\mathrm{lr}_f$) & $5 \times 10^{-6}$ & Fixed \\
2 & Learning rate classification  ($\mathrm{lr}_c$) & $\{1 \times 10^{-4}, 3 \times 10^{-4}, \mathbf{5 \times 10^{-4}}\}$  & Tuned\\
2 & Weight decay features & $1 \times 10^{-3}$ & Fixed \\
2 & Weight decay classification & $\{1 \times 10^{-4}, \mathbf{1 \times 10^{-3}}, 1 \times 10^{-2}\}$
                                                           & Tuned \\
2 & Gradient clip threshold & 1.0 & Fixed \\
2  & Batch size ($r$) & 32 & Fixed \\
\hline
\end{tabular}
\caption{Hyperparameter search space for Model~C tuned via Optuna. Optimal
  values in bold.}
\label{tab:optuna_triplet}
\end{table}

\subsubsection{Training monitoring}
\label{sec:MC-monitoring}

\begin{figure}[tbh]
  \centering
  \includegraphics[width=1\linewidth]{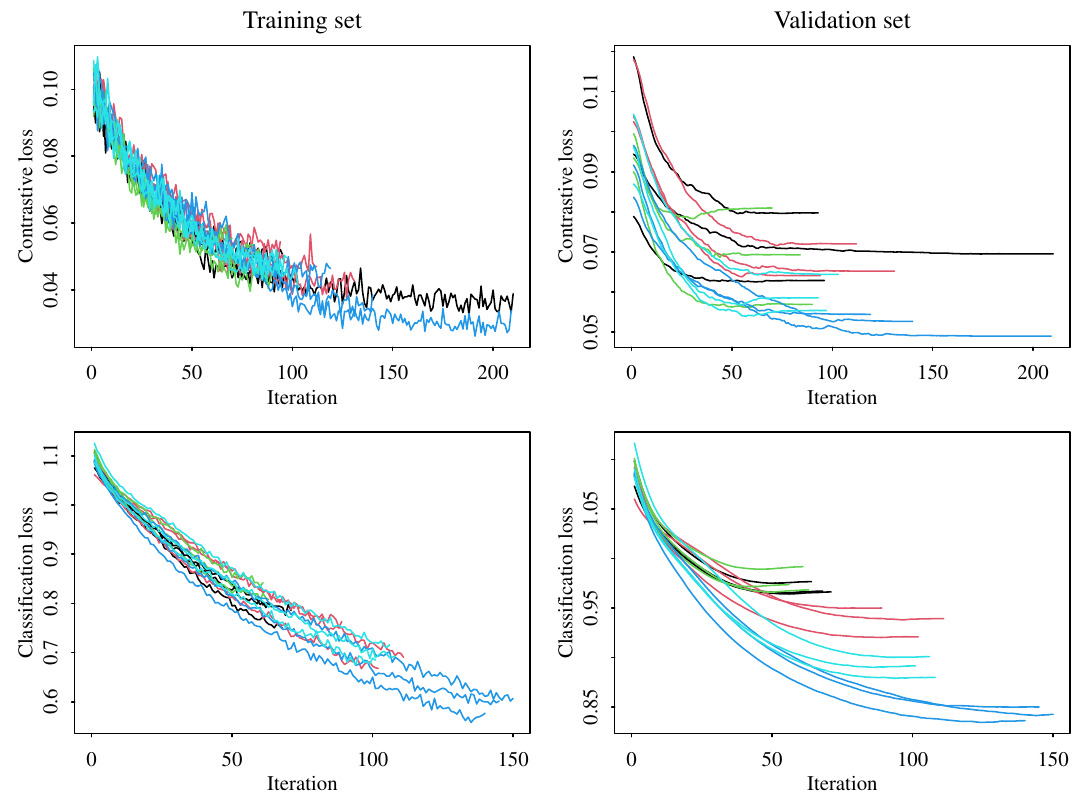}
  \caption{Training convergence curves for Model~C. The left column shows the
    contrastive and classification losses on the training set and the right column shows the
    contrastive and classification losses for the validation set against training iteration.
    Each colour corresponds to one of the data split; the three curves of the
    same colour are the three independent runs for that split. Training stops
    when the early stopping criterion is met.}
  \label{fig:ModelC_conv}
\end{figure}

Web Figure~\ref{fig:ModelC_conv} shows the classification loss for the
training and validation sets for the five different splits (different
colours) for each of the three runs for each split. Separate curves are shown
for each training stage.

\subsection{Model D: Convolutional neural network with ``word'' inputs}
\label{sec:MD}

This model consists of a convolutional neural network (CNN) model that
utilises a pre-trained language model (LM) to derive sequence embeddings from
molecular representations. The model components are as follows.
\begin{enumerate}
\item \textbf{Word tokeniser:} Each compound's SMILES string is tokenised
  using a pre-trained LM. The LM maps input strings into learnt
  components, thus creating a feature space. We consider three model variants
  from the BERT (Bidirectional Encoder Representations from
  Transformers) family \citep{devlin2018bert}:
  \begin{itemize}
  \item \textbf{BERT:} The standard BERT model (\verb|bert-base-uncased|).
  \item \textbf{DistilBERT:} A distilled version of the standard BERT (\verb|distilbert-base-uncased|).
  \item \textbf{ChemBERTa:} A BERT model pre-trained specifically on a large
    corpus of chemical text and SMILES strings (\verb|seyonec/ChemBERTa-zinc-base-v1|) \citep{chithrananda2020chemberta}.
  \end{itemize}
  Each LM generates a tensor
  $X^{(0)} \in \mathbb{R}^{r \times m_0 \times l_0}$, where $r$ denotes the
  batch size, $m_0$ is the hidden embedding dimension, and $l_0$ is the token
  sequence length. The first $2$ transformer layers of each model are loaded
  as this was found to have the best performance. 
  
\item \textbf{Hidden layers:}
  \begin{enumerate}
  \item \textbf{1D convolution:} The token output tensor $X^{(0)}$
    is processed via a 1D
    convolution with kernel size $s_1$, output channels $m_1$, and
    same-padding:
    \begin{equation*}
      Z^{(1)}_{tuv} = \sum_{i=1}^{m_0} \sum_{k=1}^{s_1} W^{(1)}_{v, i, k}
      X^{(0)}_{t, i, u + k - \lfloor s_1 / 2 \rfloor} + B^{(1)}_v,
    \end{equation*}
    $t = 1,\ldots,r$, $u = 1,\ldots, l_{0}$, and $v = 1,\ldots,m_{1}$.
  \item \textbf{Activation:} The rectified linear unit (ReLU) activation is
    applied:
    \begin{equation*}
      U^{(1)}_{tuv} = \max\{0, Z^{(1)}_{tuv}\},
    \end{equation*}
    $t = 1,\ldots,r$, $u = 1,\ldots, l_{0}$, $v = 1,\ldots,m_{1}$.
  \item \textbf{Adaptive max pooling:} The maximum across the sequence dimension
    is then computed:
    \begin{equation*}
      V^{(1)}_{tv} = \max\{U^{(1)}_{tuv},\ u= 1,\ldots, l_{0}\}.
    \end{equation*}
  \item \textbf{Dropout:} Regularisation via dropout with probability of $p_d$
    is applied to $V^{(1)}$:
    \begin{equation*}
      X^{(1)}_{tv} = V^{(1)}_{tv} \times \mathbb{1}(\Upsilon_{tv} > p_d),
    \end{equation*}
    where $\Upsilon_{tv}$ are independent uniformly distributed variables in $(0,1)$.
  \end{enumerate}
\item \textbf{Output layer:} 
  \begin{enumerate}
  \item \textbf{Linear classifier:} For $C$ target classes, logits $\tilde{Z}_t \in \mathbb{R}^C$ are derived from a fully connected layer:
    \begin{equation*}
      \tilde{Z}_{th} = \sum_{v=1}^{m_1} \tilde{W}_{hv} X^{(1)}_{tv} + \tilde{B}_h, \quad h = 1, \ldots, C.
    \end{equation*}
    with weights $\tilde{W}_{hv}$ and biases $\tilde{B}_h$, $h=1,\ldots,C$.
  \item \textbf{Softmax activation:} Class probabilities $P_t^{(h)}$ for class $h \in \{1, \ldots, C\}$ are obtained via softmax:
    \begin{equation*}
      P^{(h)}_t = \mathrm{Softmax}_h(\tilde{Z}_{t,1},\ldots,\tilde{Z}_{t,C}) = 
      \frac{\exp(\tilde{Z}_{th})}{\sum_{k=1}^C \exp(\tilde{Z}_{tk})}, \quad t = 1, \ldots, r.
    \end{equation*}
  \end{enumerate}
\item \textbf{Loss function:} We use the cross-entropy (logarithmic) loss
  \begin{equation*}
    L_{\text{class}} = -\frac{1}{r} \sum_{t=1}^r \log P_t^{(y_t)},
  \end{equation*}
  where $y_t$ denotes the observed toxicity class.

\end{enumerate}

\subsubsection{Implementation details}
\label{sec:MD-impl}

The language model underlying weights are frozen during training, leaving
only the CNN parameters $W^{(1)}, B^{(1)}$ and classifier parameters
$\tilde{W}, \tilde{B}$ to be updated. Next, we outline the specific training
protocol and hyperparameter choices. Due to the higher computational cost for
training this model, we used fewer Optuna trials to select hyperparameters.

\begin{itemize}
\item \textbf{Hyperparameter configuration:} The architectural parameters
  used are $r=16$ (batch size), $l_0 = 50$ (token sequence length),
  $m_0 = 768$ and $m_1 = 256$ (hidden dimension), $s_1 = 3$ (kernel size),
  and $p_d = 0.5$ (dropout rate).
  
\item \textbf{Optimisation protocol:} Optimised using the Adam optimiser with $L_2$ regularisation. Initial learning rate is set to $1 \times 10^{-5}$ with a weight decay penalty of $1 \times 10^{-2}$.
 Gradient norm clipping is applied at a maximum threshold of $1.0$ to ensure stability.
 A scheduler reduces the learning rate by a factor of $0.1$ when validation
 loss fails to improve for $3$ consecutive epochs.
 
\item \textbf{Training and early stopping:}
  Training uses randomly
    shuffled mini-batches of size $r = 16$.
 Models are trained  for up to $60$ epochs.
  Training halts early if the logarithmic loss
  over the validation set does not decrease by at least $0.0005$ over $5$
  consecutive epochs. Once early stopping criteria were met, the model
  weights corresponding to the epoch with the lowest validation loss were
  restored for out-of-sample test evaluation.
  
\item \textbf{Hyperparameter tuning:} Hyperparameter search space is given in
  Web Table~\ref{tab:optuna_bert_cnn}. Optuna (15 trial runs per model
  variant), based on minimising the logarithmic loss over the validation set,
  was used to select optimal values.

\item \textbf{Replications:} Each model training procedure was run 3 times
  for each split, with different parameter starting values each time 
  to ensure robustness of the training procedure.

\item \textbf{Compute Budget:} Full
  training completes in under 5 minutes per run on a single CPU.
\end{itemize}

\begin{table}[htbp]
\centering
\begin{tabular}{lll}
\hline
\textbf{Hyperparameter} & \textbf{Values} & \textbf{Status}\\
\hline
Output channels ($m_1$) & $\{64, 128, \mathbf{256}\}$  & Tuned\\
  Dropout rate ($p_d$) & $\{0.3, \mathbf{0.5}, 0.7\}$ & Tuned \\
  Kernel size ($s_1$) & 3 & Fixed \\
  Transformer layers & 2 & Fixed \\
Initial learning rate & $\{\mathbf{1 \times 10^{-5}}, 1 \times 10^{-4}, 5 \times 10^{-4}\}$ & Tuned \\
Weight decay & $\{1 \times 10^{-4}, 1 \times 10^{-3}, \mathbf{1 \times 10^{-2}}\}$ & Tuned \\
Batch size ($r$) & $\{\mathbf{16}, 32, 64\}$ & Tuned \\
Gradient clip threshold & $\{0.3, 0.5, 0.7, \mathbf{1.0}\}$ & Tuned \\ 
  Token dimension ($l_0$) & 50 & Fixed \\
\hline
\end{tabular}
\caption{Hyperparameter search space for Model~D tuned via Optuna. Optimal
  values in bold.}
\label{tab:optuna_bert_cnn}
\end{table}

\subsubsection{Training monitoring}
\label{sec:MD-monitoring}

\begin{figure}
  \centering
  \includegraphics[width=1\linewidth]{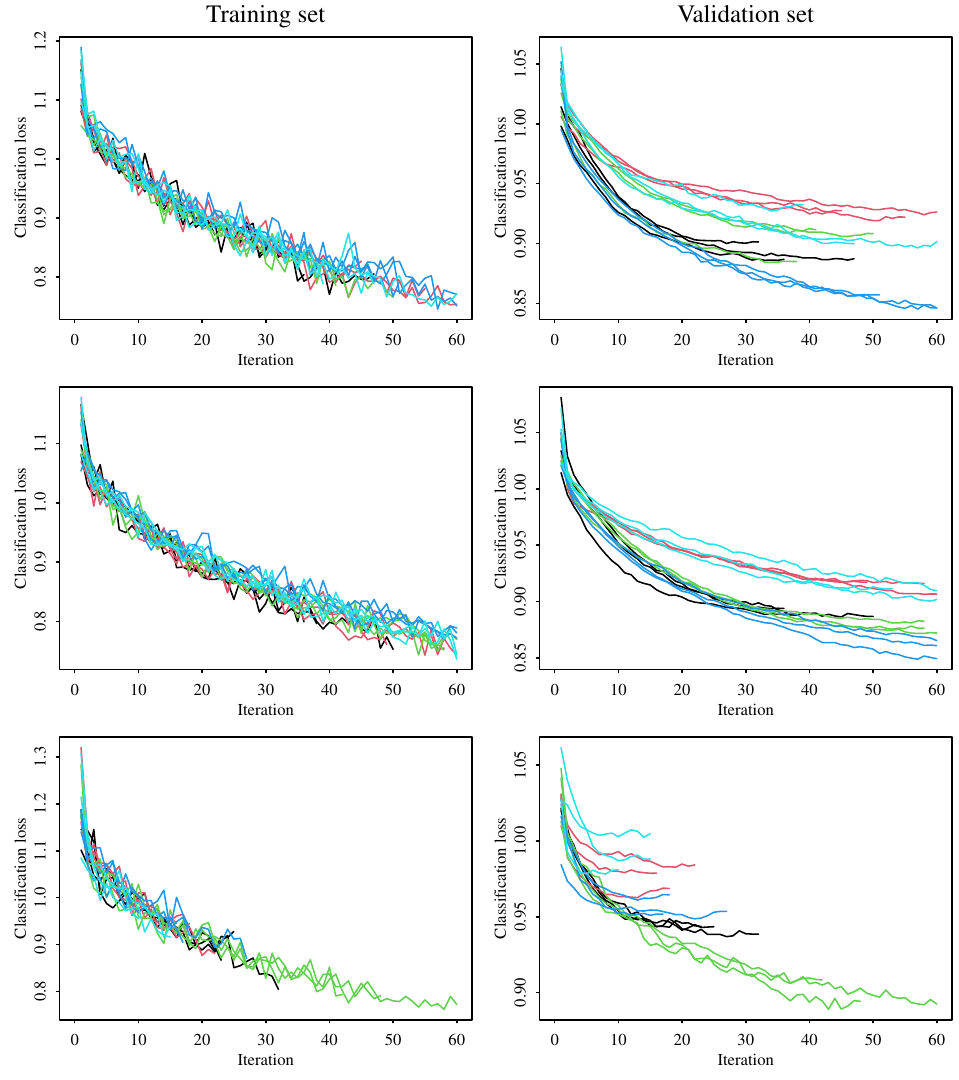}
  \caption{Training convergence curves for Model~D. First row is standard
    BERT, second row is distilled BERT, and third row is Chem BERT. The left
    column shows the classification loss on the training set and the right
    column shows the classification loss for the validation set against
    training iteration. Each colour corresponds to one of the data split; the
    three curves of the same colour are the three independent runs for that
    split. Training stops when the early stopping criterion is met.}
  \label{fig:ModelD_conv}
\end{figure}

Web Figure~\ref{fig:ModelD_conv} shows the classification loss for the
training and validation sets for the five different splits (different
colours) for each of the three runs for each split. Separate curves are shown
for each BERT variant.

\subsection{Model E: GP with Euclidean-based kernel with ``word'' inputs}
\label{sec:ME}

This model is a combination of Models~B and~D. Specifically, we extract the
embeddings $X^{(0)}_{tuv}$, $t=1,\ldots,n$, $u=1,\ldots,l_0$,
$v=1,\ldots,m_0$ from the trained language model from Model~D with the
distilled BERT model. These embeddings are used to represent the chemical
space and correspond to the inputs of a GP model with a Euclidean-based
kernel.

\subsection{Model F: Random forest model}
\label{sec:MF}

The random forest model \citep{Breiman2001} uses the eight remaining chemical
variables provided with the data (GHS classification, GHS hazard statements,
Hansen solubility parameter, boiling temperature, vapour pressure at
atmospheric pressure, density, molecular weight, and molar volume) for each
solvent instead of the chemical fingerprint. For this analysis we used the R
implementation provided by \citet{Liaw2002} with default options. For each
split, three independent training runs were executed. Each run completes in a
fraction of a second.

\subsection{Model G: XGBoost}
\label{sec:MG}

We fitted the extreme gradient boosting (XGBoost) model
\citep{chen2016xgboost} using the linear booster with a multi-class softmax
objective and with the fingerprint features as binary input variables. We
used the R implementation in \citet{xgboostR}. The learning rate was set to
0.1, with L1 and L2 regularisation parameters of 0.002 and 0.06,
respectively. Training was conducted over a maximum of 190 rounds with early
stopping patience set to 3, based on the logarithmic loss metric. For each
split, three independent training runs were executed. Each run completes in a
less than 10 seconds.

\subsection{Model H: GAUCHE + Variational Bayes}
\label{sec:MH}

GAUCHE \citep{griffiths2024gauche} is a Python library for GP for
chemoinformatics. It implements a variety of GP kernels, including, among
others, the Tanimoto kernel.

The model consists of a multi-class GP classifier. We assume $C$ independent
latent GPs $\bu^{(j)} \sim N(0,\sigma^2_j R)$, for $j=1,\ldots,C$, on the
chemical space, where $\sigma^2_j$ denotes the variance parameter and $R$
denotes a fingerprint correlation matrix. Let $\mathbf{U} = (\bu^{(1)},
\ldots, \bu^{(C)})$. For compound $c_i$, with
corresponding outcome $y_i = j$, we set
\begin{equation*}
  \Pr(y_i|\mathbf{U}(c_i)) = \mathrm{Softmax}_{j}(\mathbf{U}(c_i)) =
  \frac{\exp\{\bu^{(j)}(c_i)\}} 
  {\exp\{\bu^{(1)}(c_i)\} + \ldots + \exp\{\bu^{(C)}(c_i)\}}.
\end{equation*}
Three fingerprint correlation kernels were
considered: Tanimoto, dice, and min-max \citep[see][]{griffiths2024gauche}.

The model was trained using a variational approximation. Let $\bv^{(j)} = (\bu^{(j)}(c_1),\ldots,\bu^{(j)}(c_n))$, where
$(c_1,\ldots,c_n)$ denote the observed compounds. The posterior density
$f(\bv^{(j)}|\by)$ is approximated by a multivariate Gaussian $q_j(\bv^{(j)})$ with mean $\bm^{(j)}$ and variance
$S^{(j)} = L^{(j)} L^{(j)\top}$, where $L^{(j)}$ denotes the lower triangular
Cholesky factor. The parameters $\sigma^2_j$, $\bm^{(j)}$ and $L^{(j)}$ are estimated by
maximising the variational evidence lower bound log-likelihood (ELBO):
\begin{equation}
  \label{eq:ELBO}
  \ell(\{\sigma^2_j,\bm^{(j)},L^{(j)}: j=1,\ldots,C\}|\by) = \sum_{i=1}^n \mathbb{E}
  [\log \Pr(y_i|\mathbf{U}(c_i))] - \sum_{j=1}^C \mathrm{KL}(q_j(\bv^{(j)})\| f(\bv^{(j)})),
\end{equation}
where $\mathrm{KL}(q_j(\bv^{(j)})\| f(\bv^{(j)}))$ denotes the
Kullback-Leibler divergence of the variational approximation $q_j(\bv^{(j)})$
from the exact prior GP model, $f(\bv^{(j)})$. The expectation
$\mathbb{E} [\log \Pr(y_i|\mathbf{U}(c_i))]$ is over
$\bv^{(j)} \sim q(\bv^{(j)})$ independently for $j=1,\ldots,C$ and is
computed by Monte-Carlo using 400 independent samples from
$q(\bv^{(1)}) \times \ldots \times q(\bv^{(C)})$. Optimisation was performed
using the adaptive moment estimation (adam) optimiser \citep{kingma2015adam}.

For an unobserved compound $c_\ast$, the predicted probability for $y_\ast = j$ is calculated as
\begin{equation*}
  \Pr(y_\ast = j|\by) = \frac{1}{M} \sum_{m=1}^M
  \mathrm{Softmax}_j(\mathbf{U}_{[m]}(c_\ast)), 
\end{equation*}
where $\mathbf{U}_{[m]}(c_\ast)$ for $m=1,\ldots,M$ denote independent random
samples from the learnt posterior distribution of $\mathbf{U}(c_\ast)|\by$, with
$M=400$.

\subsubsection{Implementation details}
\label{sec:MH-impl}

The multi-class variational Gaussian Process models were implemented in
Python using GPyTorch \citep{gardner2018gpytorch} and the GAUCHE library
\citep{griffiths2024gauche}. Model parameters (kernel hyperparameters,
variational distribution parameters) were trained jointly by minimising the
negative variational ELBO \eqref{eq:ELBO}.

\begin{itemize}
\item \textbf{Inducing Strategy:} The variational distribution uses the
  complete training set ($N = n$) as fixed inducing points as the data
  set is small.
  
\item \textbf{Likelihood and Monte Carlo Sampling:} Softmax likelihood
  evaluation and prediction use $M = 400$ independent Monte-Carlo samples per
  forward pass.
  
\item \textbf{Optimisation protocol:} Model and likelihood parameters were
  trained using the Adam optimiser with an initial learning rate of
  $\eta = 0.05$ across all three kernel choices. 

\item \textbf{Training and early stopping:} Training was run for up to
  $400$ epochs, with early stopping triggered if the logarithmic loss over
  the validation set failed to improve by at least $1 \times 10^{-4}$ over
  $30$ consecutive epochs. Once early stopping criteria were met, the model
  weights corresponding to the epoch with the lowest validation loss were
  restored for out-of-sample test evaluation.

\item \textbf{Replications:} Each model training procedure was run 3 times
  for each split, with different parameter starting values each time 
  to ensure robustness of the training procedure.

\item \textbf{Compute Budget:} Full
  training completes in under a minute per run on a single CPU.
\end{itemize}

\subsubsection{Training monitoring}
\label{sec:MH-monitoring}

\begin{figure}
  \centering
  \includegraphics[width=1\linewidth]{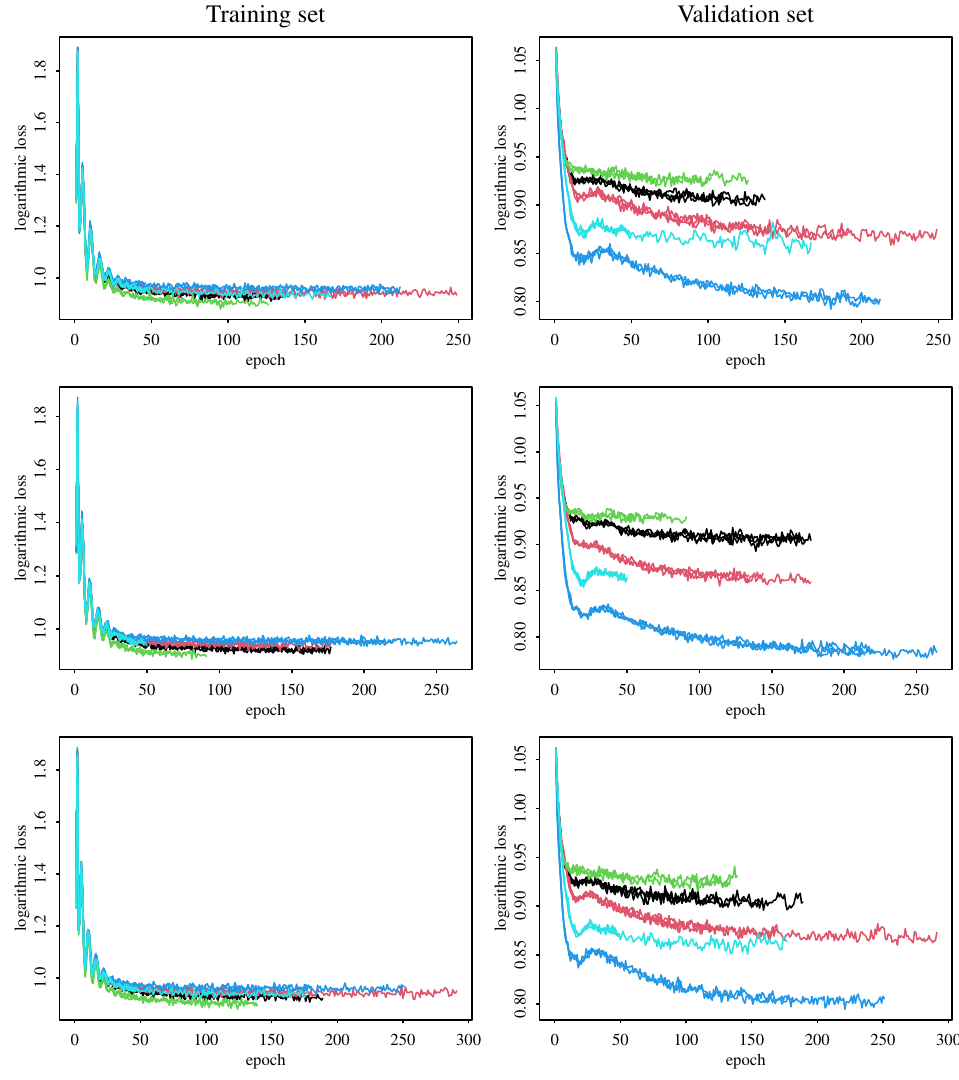}
  \caption{Training convergence curves for Model~H. First row is standard
    Tanimoto, second row is Dice, and third row is MinMax kernel. The left
    column shows the classification loss on the training set and the right
    column shows the classification loss for the validation set against
    training iteration. Each colour corresponds to one of the data split; the
    three curves of the same colour are the three independent runs for that
    split. Training stops when the early stopping criterion is met.}
  \label{fig:ModelH_conv}
\end{figure}

Web Figure~\ref{fig:ModelH_conv} shows the classification loss for the
training and validation sets for the five different splits (different
colours) for each of the three runs for each split. Separate curves are shown
for each kernel variant.

\subsection{Model I: Generalised linear mixed model}
\label{sec:MI}

This model is a variant of Model~A, but with independent covariance. The
model does not make use of the fingerprint data and is used as a benchmark
model.

\subsection{Results}
\label{sec:results}

All models were evaluated against how well they predict the class of the
solvents in the test set against three scoring rules. Let $y_* \in \{1,2,3\}$
denote the WGK classification of an arbitrary solvent in the test set, and
let $\mathbf{P}_* = (P_*(1), P_*(2), P_*(3))$ denote the estimated
probabilities for each class under the assessed model for that solvent. We
define the following losses (lower values indicate a better model).
\begin{itemize}
\item \textbf{Logarithmic loss:} $L_\text{log}(\mathbf{P}_*, y_*) = -\log
  P_*(y_*)$. 
\item \textbf{Spherical loss:}
  $L_\text{sph}(\mathbf{P}_*, y_*) = 1-\dfrac{P_*(y_*)}{\sqrt{\sum_{h=1}^3
      P_*(h)^2}}$.
\item \textbf{Misclassification:}
  $L_\text{mc}(\mathbf{P}_*, y_*) = 1 - \mathbb{1}(y_* = \argmax_{h}
  P_*(h))$. 
\end{itemize}

As the results vary by split, to aggregate the results across different
splits, we consider the difference in the average loss across different runs
for each model minus the loss for a reference model, which in this case we
take it to be Model~A with the logit link and Tanimoto correlation. Web
Table~\ref{tab:loss_results} shows the average loss difference, the standard
deviation of the loss difference, and the model rank for each loss criterion
and for each model across all data in the test set, as well as each model's
computing time.

We observe that the GP model with the
Tanimoto-based kernel (Model~A) has overall the best performance, with the
Tanimoto correlation and either the logit or the probit link ranking first
under all three loss measures.
The GP models based on the GAUCHE kernels
(Model~H) come next, followed by the GP model whose kernel is derived from
the CNN embeddings (Model~B) and the model that uses a CNN with a
pre-trained language model (Model~D).
The differences between these models
are small in comparison with the variation between data splits, so we do not
read too much into their exact ordering.
Based on these results, we can see
that the models which account for the similarity between compounds through a
GP perform at least as well as the more advanced pre-trained models, and that
Model~A can be a suitable model for chemoinformatics.
In contrast, the
generalised linear mixed model (Model~I), which makes no use of the chemical
representation, is the worst model under all three criteria, which confirms
that the fingerprints carry useful information about the hazard class.
Furthermore, looking into model specifics, we can see that for Models~A and~B the
complementary log-log (c-log-log) link function performs worse than
the other three link functions used, although the same is not true for
Model~E. Within Model~A, the Tanimoto correlation gives a lower loss than the
Gaussian correlation, which in turn gives a lower loss than the exponential
correlation. This shows that a careful model choice and assessment is
imperative for accurate predictions.

\begin{table}
  \centering
  \small\hspace*{-1cm}
\begin{tabular}{llrrrrrrrrrr}
\toprule
 & & \multicolumn{3}{c}{Log loss} & \multicolumn{3}{c}{Spherical loss} & \multicolumn{3}{c}{Misclassification} & \\
Model & Specifics & Mean & SD & Rank & Mean & SD & Rank & Mean & SD & Rank & Time (s)\\
\midrule \multirow{12}{*}{A}
 & logit exponential & 0.016 & 0.008 & 18 & 0.006 & 0.003 & 14 & 0.000 & 0.022 & 1 & 6.4\\
 & probit exponential & 0.011 & 0.006 & 12 & 0.004 & 0.002 & 11 & 0.003 & 0.018 & 7 & 6.5\\
 & log-log exponential & 0.019 & 0.003 & 20 & 0.009 & 0.002 & 19 & 0.023 & 0.025 & 30 & 5.4\\
 & c-log-log exponential & 0.044 & 0.019 & 33 & 0.017 & 0.008 & 32 & 0.006 & 0.013 & 9 & 7.0\\
\addlinespace
 & logit Gaussian & 0.002 & 0.002 & 5 & 0.001 & 0.001 & 5 & 0.000 & 0.000 & 1 & 10.0\\
 & probit Gaussian & 0.001 & 0.005 & 3 & 0.001 & 0.002 & 4 & 0.003 & 0.012 & 8 & 15.2\\
 & log-log Gaussian & 0.006 & 0.010 & 9 & 0.002 & 0.003 & 8 & 0.008 & 0.016 & 13 & 9.6\\
 & c-log-log Gaussian & 0.019 & 0.011 & 19 & 0.006 & 0.004 & 15 & 0.000 & 0.017 & 1 & 12.5\\
\addlinespace
 & logit Tanimoto & 0.000 & NA & 1 & 0.000 & NA & 1 & 0.000 & NA & 1 & 6.7\\
 & probit Tanimoto & 0.000 & 0.004 & 1 & 0.000 & 0.001 & 2 & 0.000 & 0.010 & 1 & 5.9\\
 & log-log Tanimoto & 0.004 & 0.011 & 6 & 0.001 & 0.004 & 6 & 0.011 & 0.015 & 16 & 4.9\\
 & c-log-log Tanimoto & 0.009 & 0.009 & 10 & 0.002 & 0.003 & 10 & 0.000 & 0.017 & 1 & 6.5\\
\midrule \multirow{8}{*}{B}
 & logit exponential & 0.013 & 0.011 & 13 & 0.006 & 0.005 & 13 & 0.006 & 0.017 & 9 & 12.1\\
 & probit exponential & 0.010 & 0.009 & 11 & 0.005 & 0.005 & 12 & 0.006 & 0.019 & 9 & 11.0\\
 & log-log exponential & 0.015 & 0.006 & 16 & 0.008 & 0.004 & 16 & 0.017 & 0.018 & 24 & 9.2\\
 & c-log-log exponential & 0.037 & 0.020 & 31 & 0.015 & 0.009 & 30 & 0.021 & 0.019 & 29 & 9.3\\
\addlinespace
 & logit Gaussian & 0.022 & 0.018 & 23 & 0.010 & 0.009 & 21 & 0.011 & 0.018 & 18 & 33.8\\
 & probit Gaussian & 0.020 & 0.021 & 21 & 0.010 & 0.009 & 20 & 0.011 & 0.010 & 17 & 27.2\\
 & log-log Gaussian & 0.022 & 0.015 & 24 & 0.010 & 0.007 & 21 & 0.016 & 0.022 & 21 & 26.6\\
 & c-log-log Gaussian & 0.033 & 0.023 & 30 & 0.015 & 0.010 & 29 & 0.018 & 0.022 & 26 & 29.6\\
\midrule \multirow{2}{*}{C}
 & contrastive learning & 0.016 & 0.025 & 17 & 0.009 & 0.012 & 17 & 0.016 & 0.022 & 22 & 205.8\\
 & benchmark & 0.040 & 0.046 & 32 & 0.020 & 0.022 & 35 & 0.045 & 0.041 & 38 & 5.0\\
\midrule \multirow{3}{*}{D}
 & standard BERT & 0.029 & 0.022 & 28 & 0.016 & 0.013 & 31 & 0.030 & 0.054 & 32 & 249.5\\
 & distilled BERT & 0.014 & 0.028 & 14 & 0.011 & 0.016 & 24 & 0.017 & 0.067 & 25 & 249.5\\
 & chem-BERT & 0.058 & 0.047 & 37 & 0.027 & 0.025 & 38 & 0.041 & 0.052 & 35 & 249.5\\
\midrule \multirow{8}{*}{E}
 & logit exponential & 0.023 & 0.028 & 25 & 0.011 & 0.015 & 25 & 0.013 & 0.056 & 19 & 256.0\\
 & probit exponential & 0.026 & 0.027 & 26 & 0.012 & 0.015 & 27 & 0.009 & 0.060 & 14 & 257.9\\
 & log-log exponential & 0.030 & 0.027 & 29 & 0.011 & 0.014 & 23 & 0.010 & 0.045 & 15 & 256.2\\
 & c-log-log exponential & 0.015 & 0.032 & 15 & 0.009 & 0.017 & 18 & 0.007 & 0.059 & 12 & 254.7\\
\addlinespace
 & logit Gaussian & 0.050 & 0.025 & 34 & 0.020 & 0.015 & 34 & 0.042 & 0.061 & 36 & 272.5\\
 & probit Gaussian & 0.057 & 0.025 & 35 & 0.021 & 0.015 & 36 & 0.032 & 0.070 & 33 & 266.6\\
 & log-log Gaussian & 0.068 & 0.032 & 38 & 0.019 & 0.013 & 33 & 0.026 & 0.059 & 31 & 268.2\\
 & c-log-log Gaussian & 0.028 & 0.026 & 27 & 0.015 & 0.016 & 28 & 0.021 & 0.051 & 28 & 270.8\\
\midrule \multirow{1}{*}{F}
 & Random Forest & 0.058 & 0.037 & 36 & 0.023 & 0.009 & 37 & 0.037 & 0.050 & 34 & 0.2\\
\midrule \multirow{1}{*}{G}
 & XGBoost & 0.021 & 0.029 & 22 & 0.011 & 0.014 & 26 & 0.044 & 0.032 & 37 & 7.6\\
\midrule \multirow{3}{*}{H}
 & Tanimoto & 0.005 & 0.018 & 7 & 0.002 & 0.008 & 7 & 0.016 & 0.012 & 23 & 27.4\\
 & Dice & 0.002 & 0.018 & 4 & 0.001 & 0.009 & 3 & 0.015 & 0.013 & 20 & 22.5\\
 & MinMax & 0.006 & 0.015 & 8 & 0.002 & 0.006 & 9 & 0.020 & 0.010 & 27 & 76.7\\
\midrule \multirow{4}{*}{I}
 & logit & 0.125 & 0.052 & 42 & 0.063 & 0.027 & 42 & 0.090 & 0.067 & 42 & 1.9\\
 & probit & 0.123 & 0.051 & 39 & 0.061 & 0.026 & 39 & 0.085 & 0.063 & 39 & 2.5\\
 & log-log & 0.125 & 0.051 & 41 & 0.063 & 0.026 & 41 & 0.085 & 0.063 & 39 & 2.2\\
 & c-log-log & 0.124 & 0.051 & 40 & 0.062 & 0.026 & 40 & 0.088 & 0.066 & 41 & 4.8\\
\bottomrule
\end{tabular}
\caption{Average classification loss difference compared to Model~A with
  logit link and Tanimoto correlation for each model across the test data.
  The standard deviation of the difference and the model rank (from lowest to
  highest average loss) are also shown for three different loss measures.
  Model computing time in seconds is also shown.}
  \label{tab:loss_results}
\end{table}

\begin{figure}
  \centering
  \includegraphics[width=1\linewidth]{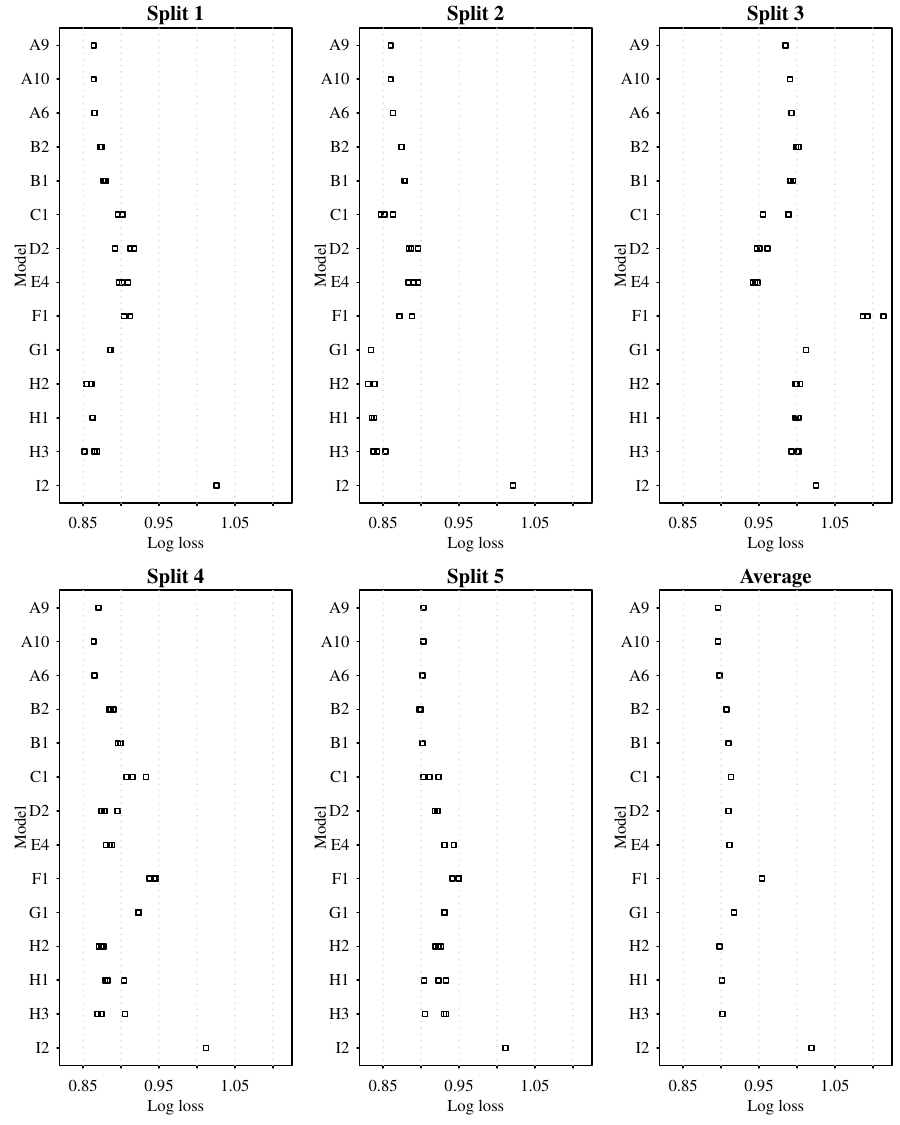}
  \caption{Individual logarithmic loss values for each independent run
    plotted separately for each data split and average over all splits for
    selected models. The models are as appear in Table~\ref{tab:solvents1}.}
  \label{fig:log_loss_strip}
\end{figure}

Web Figure~\ref{fig:log_loss_strip} shows the individual logarithmic loss
values for each separate run for the selection of models shown in
Table~\ref{tab:solvents1}. We can observe that within each data split, there
is little variability between model runs, however, across different splits,
the results vary significantly. The test subset for Split~3 differs the most
from the training and validation subsets, and is also the hardest to predict. Nevertheless, we can observe that the
proposed Model~A has consistently good performance across splits even if it
is not always the best model.

\printbibliography[title={Supplementary References}]

\end{refsection}

\end{document}